\pdfoutput=1
\documentclass[cernpreprint, atlasdraft=false, texlive=2025, UKenglish, texmf, orcidlogo]{atlasdoc}
\usepackage{atlaspackage}
\usepackage{atlasbiblatex}

\usepackage{atlascontribute}

\usepackage{atlasphysics}
\graphicspath{{logos/}{figures/}}

\usepackage{ANA-TOPQ-2018-23-PAPER-defs}
\usepackage{siunitx}
\usepackage{xcolor}
\usepackage{multirow}
\usepackage{comment}
\usepackage{blindtext}
\usepackage{rotating}
\definecolor{amarillo}{rgb}{255,255,0}
\definecolor{bananamania}{rgb}{0.98, 0.91, 0.71}
\definecolor{celadon}{rgb}{0.67, 0.88, 0.69}
\definecolor{salmonpink}{rgb}{1.0, 0.57, 0.64}

\AtlasTitle{Measurement of the top-quark pole mass in dileptonic $t\bar{t}+ 1\text{-jet}$ events at \ensuremath{\sqrt{s}=13}~TeV with the ATLAS experiment}

\AtlasJournalRef{\JHEP 12 (2025) 023}
\AtlasDOI{10.1007/JHEP12(2025)023}
\AtlasRefCode{TOPQ-2018-23}
\PreprintIdNumber{CERN-EP-2025-135}

\AtlasAbstract{
A measurement of the top-quark pole mass $m_{t}^\text{pole}$ is presented  in $t\bar{t}$~events with an additional jet, $t\bar{t}+1\text{-jet}$, produced in $pp$ collisions at $\sqrt{s}=13$~TeV.
The data sample, recorded with the ATLAS experiment during Run~2 of the LHC, corresponds to an integrated luminosity of $140~\text{fb}^{-1}$.
Events with one electron and one muon of opposite electric charge in the final state are selected to measure the $t\bar{t}+1\text{-jet}$  differential cross-section as a function of the inverse of the invariant mass of the $t\bar{t}+1\text{-jet}$ system. Iterative Bayesian Unfolding is used to correct the data to enable comparison with fixed-order calculations at next-to-leading-order accuracy in the strong coupling.
The process $pp \to t\bar{t}j$ ($2 \rightarrow 3$), where top quarks are taken as stable particles, and the  process $pp \to b\bar{b}l^+\nu l^- \bar{\nu} j$ ($2 \to 7$), which includes top-quark decays to the dilepton final state and off-shell effects, are considered.
The top-quark mass is extracted using a $\chi^2$ fit of the unfolded normalized differential cross-section distribution. The results obtained with the $2 \to 3$ and $2 \to 7$ calculations are compatible within theoretical uncertainties, providing an important consistency check. %
The more precise determination is obtained for the $2 \to 3 $ measurement: %
$m_{t}^\text{pole}=170.7\pm0.3~(\text{stat.})\pm1.4~(\text{syst.})~\pm 0.3~(\text{scale})~\pm 0.2~(\text{PDF}\oplus\alpha_\text{S})~\text{GeV},$
which is in good agreement with other top-quark mass results.
}

\author{The ATLAS Collaboration}

\usepackage{authblk}
\AtlasRefCode{TOPQ-2018-23}

\AtlasNote{TOPQ-2018-23}

\hypersetup{pdftitle={ATLAS document},pdfauthor={The ATLAS Collaboration}}

\begin{document}

\maketitle

\tableofcontents

\clearpage

\section{Introduction}
\label{sec:intro}

% The next lines are included from the .//content/Introduction.tex input file
The mass of the top quark plays a fundamental role in the Standard Model (SM) of particle physics.
The parameters of the SM Lagrangian --- the masses of the elementary particles and the strength of their interactions --- are not predicted by the SM, but must be measured experimentally.
In the SM, the masses of the elementary particles arise from interactions with the Higgs field.
The top quark Yukawa coupling is approximately 1 and the top quark mass strongly affects the evolution of the Higgs quartic coupling to high scales~\cite{Alekhin:2012py,Degrassi:2012ry}, which determines the shape of the Higgs potential and the stability of the quantum vacuum.
The SM also predicts a relation among the masses of the Higgs boson, the $W$ boson~\cite{LHC-TeVMWWorkingGroup:2023zkn}, and the top quark.
Precise measurements of these masses and other fundamental parameters allow a test of this important relation~\cite{Haller2018,PhysRevLett.129.271.801k}.

Direct measurements of the top-quark mass parameter in Monte Carlo (MC) generators, based on fits to the observed distributions of top-quark decay products, have reached excellent experimental precision~\cite{ATLAS:2024dxp}.
The results of these measurements are generally identified with the top-quark pole mass. The ambiguity in this identification, from limitations of the stable-top-quark picture and non-perturbative effects, is estimated to be a few hundred \MeV~\cite{Azzi:2019yne}. Work is ongoing to improve the understanding of the top-quark mass parameter in Monte Carlo generators~\cite{Hoang:2020iah,FerrarioRavasio2018}. The pole mass definition moreover has an intrinsic renormalon ambiguity~\cite{Beneke:2021lkq, Beneke:2016cbu}.

The top-quark mass can also be extracted from measurements of the top-quark pair production cross-section, either through inclusive or differential approaches. In this case, the mass scheme is defined by the fixed-order (FO) prediction for the cross-section in quantum chromo dynamics (QCD). This leaves no ambiguity when such measurements are used in other phenomenological studies done in perturbation theory, such as the calculations of the stability of the electroweak (EW) vacuum. The mass extraction procedure yields an estimate of the theory uncertainty, greater flexibility in the choice of the mass scheme, and the possibility to take advantage of future theory progress without redoing the measurement.

In this paper, the top-quark pole mass \Mpole\ is measured in \ttbar\ events with an additional jet (\ttbarjet) produced in proton--proton (\pp) collisions at $\sqrt{s} =$ 13~\TeV.
Following the method proposed in Ref.~\cite{Alioli:2013mxa}, the dependence of the normalized cross-section of \ttbarjet\ production as a function of the invariant mass $\sqrt{s_{\ttbarjet~}}$ of the \ttbarjet~ system on \Mpole\ is studied.
In the differential cross section, the top quark pair production threshold region at $m_{t\bar t} \approx 2 m_t$ has a pronounced top-quark mass sensitivity which is further enhanced in the \ttbarjet~final state, as the additional jet allows events from the threshold region to be experimentally more accessible.
At the same time the impact of a bound-state enhancement at threshold~\cite{Ju2020,Garzelli:2024uhe,Kiyo2009,Sumino2010,HAGIWARA200871,Nason:2025hix,CMS:2025kzt,Fadin1990} not included in fixed-order calculations, is diluted by the presence of the additional jet.
The observable is defined as:
\begin{equation}\label{eq:R_definition}
\mathcal{R}(\Rhos; m^{\text{pole}}_t) = \frac{1}{\sigma_{\ttbarjet~}} \cdot \frac{\text{d}\sigma_{\ttbarjet~}}{\text{d}\Rhos}\quad \text{, with }\quad  \Rhos = \frac{2 m_0}{\sqrt{s_{\ttbarjet~}}}.
\end{equation}
The constant $m_0$ is set to $m_0=170$~\GeV~to ensure that \Rhos~is bounded between 0 and 1, as in Ref.~\cite{TOPQ-2014-06}. This convention is arbitrary and has no impact on the final top-quark mass measurement.

Previously, the ATLAS collaboration measured the top-quark pole mass in \ttbarjet\ events using $\sqrt{s}=7~\TeV$~\cite{TOPQ-2014-06} and $\sqrt{s}=8~\TeV$~\cite{TOPQ-2019-12} $pp$ collision data, in the semileptonic \ttbar~decay channel, resulting in total uncertainties of $2.3~\GeV$ (1.3\%) and $1.2~\GeV$ (0.7\%) on the top-quark pole mass, respectively.
This analysis targets the dilepton decay channel of the \ttbar\ system and uses the complete Run~2 data sample, corresponding to 140~\ifb, allowing the dilepton channel to provide competitive sensitivity.
The CMS collaboration published the first measurement of the top-quark pole mass in \ttbarjet\ events in the dilepton channel~\cite{CMS-TOP-21-008} at $\sqrt{s}=13~\TeV$ using \lumiCMSSixteen\ resulting in %
a total uncertainty of $1.36~\GeV$ (0.8\%).

The final state targeted by this analysis has two charged leptons, two neutrinos, two $b$-jets, and an additional jet with high transverse momentum. Events are selected that have an electron and a muon candidate, at least three jets and a significant amount of missing transverse momentum. The four-momentum of the \ttbar system is reconstructed using a combination of two methods, as detailed in Section~\ref{sec:ObjectsAndEventSelection}.

The $\rho_s$ distribution is  corrected to the level of stable top quarks, undoing the impact of top-quark decay, hadronization, and detector response, using Iterative Bayesian Unfolding (IBU)~\cite{DAgostini:1994fjx}.
The top-quark pole mass is  obtained from a fit of the normalized differential cross-section \Robs\ to a FO calculation of the $2\rightarrow3$ process $pp\rightarrow \ttbarjet$ at next-to-leading-order (NLO) accuracy in QCD~\cite{Alioli:2011as}.

The data are also corrected to a parton level phase space defined by leptons, neutrinos and jets formed of quarks and gluons just before hadronization. This enables a comparison of the measured observable with a FO NLO QCD calculation for the $2 \to 7$ process $pp \to b\bar{b} l^+ \nu l^- \bar{\nu} j$ that includes top-quark decay effects and full non-resonant contributions~\cite{Bevilacqua:2016jfk}. In this second approach there is no need to correct the data for the top-quark decay, and hence a smaller reliance on the description of the decay in Monte Carlo generators. Associated single top-quark production with a $W$ boson is included in the $2 \to 7$ calculation.

This paper is structured as follows.
A brief description of the ATLAS detector is found in Section~\ref{sec:introATLAS}.
The data and MC samples used in this analysis are summarized in Section~\ref{sec:Analysis:sub:DataAndMC}.
The object definitions, object selections, and control distributions are shown in Section~\ref{sec:ObjectsAndEventSelection}.
The parton-level definitions and the FO predictions are described in Section~\ref{sec:PartonLevelAndFixedOrder}.
The reconstruction of the \ttbarjet\ system and the unfolding of the differential cross-section are described in Section~\ref{sec:DifferentialCrossSection}, and the estimate of the experimental uncertainties affecting the measurement is outlined  in Section~\ref{sec:SystematicUncertainties}.
The extraction of the top-quark mass in the pole mass scheme using a fit to the FO \ttbarjet\ predictions is discussed in Section~\ref{sec:PoleMassFit}, together with a description of the theoretical uncertainties associated with the mass measurement.
The results are then summarized in Section~\ref{sec:Results} and  conclusions are given in Section~\ref{sec:Conclusion}.

% End of text imported from the .//content/Introduction.tex input file

%
\section{ATLAS detector}
\label{sec:introATLAS}

%

% The next lines are included from the .//content/ATLASdetector_long.tex input file
\newcommand{\AtlasCoordFootnote}{%
ATLAS uses a right-handed coordinate system with its origin at the nominal interaction point (IP)
in the center of the detector and the \(z\)-axis along the beam pipe.
The \(x\)-axis points from the IP to the center of the LHC ring,
and the \(y\)-axis points upwards.
Cylindrical coordinates \((r,\phi)\) are used in the transverse plane,
\(\phi\) being the azimuthal angle around the \(z\)-axis.
The pseudorapidity is defined in terms of the polar angle \(\theta\) as \(\eta = -\ln \tan(\theta/2)\).
Angular distance is measured in units of \(\Delta R \equiv \sqrt{(\Delta\eta)^{2} + (\Delta\phi)^{2}}\).}

The ATLAS detector~\cite{PERF-2007-01} at the LHC covers nearly the entire solid angle around the collision point.\footnote{\AtlasCoordFootnote}
It consists of an inner tracking detector surrounded by a thin superconducting solenoid, electromagnetic and hadronic calorimeters,
and a muon spectrometer incorporating three large superconducting air-core toroidal magnets.

The inner-detector system (ID) is immersed in a \qty{2}{\tesla} axial magnetic field
and provides charged-particle tracking in the range \(|\eta| < 2.5\).
The high-granularity silicon pixel detector covers the vertex region and typically provides four measurements per track,
the first hit generally being in the insertable B-layer (IBL) installed before Run~2~\cite{ATLAS-TDR-19,PIX-2018-001}.
It is followed by the SemiConductor Tracker (SCT), which usually provides eight measurements per track.
These silicon detectors are complemented by the transition radiation tracker (TRT),
which enables radially extended track reconstruction up to \(|\eta| = 2.0\).
The TRT also provides electron identification information
based on the fraction of hits (typically 30 in total) above a higher energy-deposit threshold corresponding to transition radiation.

The calorimeter system covers the pseudorapidity range \(|\eta| < 4.9\).
Within the region \(|\eta|< 3.2\), electromagnetic calorimetry is provided by barrel and
endcap high-granularity lead/liquid-argon (LAr) calorimeters,
with an additional thin LAr presampler covering \(|\eta| < 1.8\)
to correct for energy loss in material upstream of the calorimeters.
Hadronic calorimetry is provided by the steel/scintillator-tile calorimeter,
segmented into three barrel structures within \(|\eta| < 1.7\), and two copper/LAr hadronic endcap calorimeters.
The solid angle coverage is completed with forward copper/LAr and tungsten/LAr calorimeter modules
optimized for electromagnetic and hadronic energy measurements, respectively.

The muon spectrometer (MS) comprises separate trigger and
high-precision tracking chambers measuring the deflection of muons in a magnetic field generated by the superconducting air-core toroidal magnets.
The field integral of the toroids ranges between \num{2.0} and \qty{6.0}{\tesla\metre}
across most of the detector.
Three layers of precision chambers, each consisting of layers of monitored drift tubes, cover the region \(|\eta| < 2.7\),
complemented by cathode-strip chambers in the forward region, where the background is highest.
The muon trigger system covers the range \(|\eta| < 2.4\) with resistive-plate chambers in the barrel, and thin-gap chambers in the endcap regions.

The luminosity is measured mainly by the LUCID--2~\cite{LUCID2} detector that records Cherenkov light produced in the quartz windows of photomultipliers located close to the beampipe.

Events were selected by the first-level trigger system implemented in custom hardware,
followed by selections made by algorithms implemented in software in the high-level trigger~\cite{TRIG-2016-01}.
The first-level trigger accepted events from the \qty{40}{\MHz} bunch crossings at a rate close to \qty{100}{\kHz},
which the high-level trigger further reduced in order to record complete events to disk at about \qty{1.25}{\kHz}.

A software suite~\cite{SOFT-2022-02} is used in data simulation, in the reconstruction
and analysis of real and simulated data, in detector operations, and in the trigger and data acquisition
systems of the experiment.

% End of text imported from the .//content/ATLASdetector_long.tex input file

%
%
%

% The next lines are included from the .//content/Analysis.tex input file
%
\section{Data and simulated samples}
\label{sec:Analysis:sub:DataAndMC}
%

% The next lines are included from the .//content/Analysis-DataMC.tex input file

This analysis is performed using the full Run 2 \pp\ data recorded with the ATLAS experiment at $\sqrt{s} = 13$~\TeV.
Once all quality requirements~\cite{DAPR-2018-01} are applied, the recorded data corresponds to an integrated luminosity of $140.0\pm 1.2~\ifb$~\cite{DAPR-2021-01}, obtained using the LUCID-2 detector~\cite{LUCID2} for the primary luminosity measurements, complemented by measurements using the inner detector and calorimeters.

The events in this data sample are required to satisfy the trigger requirements for single electrons or muons~\cite{TRIG-2018-01,TRIG-2018-05,TRIG-2019-03,TRIG-2016-01}.
%

% End of text imported from the .//content/Analysis-DataMC.tex input file

%
%

% The next lines are included from the .//content/Analysis-DataMC-Signal.tex input file
Simulated MC events are used to correct the data and estimate background processes.
Detector effects are simulated fully using \GEANT~\cite{GEANT4:2002zbu}. A few MC samples used for tests and crosschecks employ a faster detector simulation (\texttt{AtlFast2}) that makes use of parameterized showers in the calorimeters~\cite{SOFT-2010-01}.

The nominal \ttbar\ MC sample is generated using \POWHEGBOXV{2}%
~\cite{Nason:2004rx,Frixione:2007vw,Alioli:2010xd} with the heavy-quark production (\textsc{hvq}) process at NLO accuracy in QCD~\cite{Frixione:2007nw}, assuming a top-quark mass of $\MtopMC = \TopMassNominal$~\gev~and dileptonic top-quark pair decay.
The proton structure in the Matrix Element (ME) is modeled using the  NNPDF 3.0 NLO parton distribution functions (PDF) set~\cite{Ball:2014uwa}.
The \POWHEGBOXV{2} event generator is interfaced with \PYTHIA[8.230]~\cite{Sjostrand:2014zea} which simulates parton shower (PS), fragmentation, hadronization, and the underlying event (UE).
The $\hdamp$ parameter that controls the emission of the first gluon is set to $1.5\cdot \MtopMC$, while the factorization scale $\muR$ and renormalization scale $\muF$ are set to $\muR^2 = \muF^2 = (\Mtop^2 + \pT^2 )$, where the top-quark transverse momentum (\pT) is evaluated before radiation~\cite{ATL-PHYS-PUB-2016-020}.
The A14 ATLAS UE tune~\cite{ATL-PHYS-PUB-2014-021}  together with the NNPDF 2.3 leading-order (LO) PDF set~\cite{Ball_2013} is applied for \PYTHIAV{8} showering.
Finite top-quark width effects are implemented by reweighting the simulated events such that the top-quark mass distribution follows a Breit--Wigner distribution.
The inclusive \ttbar\ cross-section in the simulation is normalized to the next-to-next-to-leading-order (NNLO) cross-section including the resummation of soft gluon emissions at next-to-next-to-leading-logarithmic (NNLL) accuracy calculated using the \textsc{Top++2.0} software~\cite{Czakon:2014}.
The resulting cross-section for the \ttbar~process is $\sigma_{\ttbar} = \TTbarCrossSec$~pb~\cite{Beneke:2011mq,Cacciari:2011hy,Barnreuther:2012wtj,Czakon:2012zr,Czakon:2012pz,Czakon:2013goa}.
To estimate the systematic uncertainty associated with the \ttbar~nominal sample, additional simulations are generated, as detailed in Section~\ref{sec:Results:sub:SystematicUncertainties:subsub:Backgroundmodelling}.
\begin{comment} %
To estimate the systematic uncertainty due to the choice of the parton shower and hadronization models, the \POWHEG\ simulation is interfaced with \HERWIGV{7.1.3}~\cite{Bellm:2015jjp} using the NNPDF 3.0 NLO  PDF set in the ME, and the MMHT 2014 LO~\cite{Harland-Lang:2014zoa} one in the parton shower and the multi-parton interaction model.
Modeling uncertainties in the description of initial and final state radiation and due to the choice of renormalization and factorization scales are evaluated by internal reweighting of the \POWPYTHIA~8 sample. Uncertainties related to the shower and hadronization, matrix-element matching, color reconnection, underlying events, recoil effects in top-quark decays, and the $\hdamp$ parameter in \POWHEG, are estimated by comparing the benchmark simulated data set with samples where the corresponding parameters are varied, as detailed in Section~\ref{sec:Results:sub:SystematicUncertainties:subsub:Backgroundmodelling}.
Variations of the PDF and the strong coupling constant $\alphas$ selections are assessed in accordance with the PDF4LHC prescriptions~\cite{PDF4LHC2016}.
\end{comment}

The \singletW\  production process is simulated using \POWHEGBOXV{2} interfaced with \PYTHIAV{8} with the A14 UE tune and the NNPDF2.3 LO PDF set. Samples are normalized to the cross-section calculation at approximate NNLO accuracy~\cite{Kidonakis:2021vob}.
The overlap between $\ttbar$ and \singletW\ production is removed using the diagram-removal procedure~\cite{Kidonakis:2010}, and an additional \singletW\ simulation employing the alternative diagram-subtraction scheme~\cite{White:2009yt} is used to study its associated uncertainty.

Further \ttbar\ and \singletW\ samples with varied top-quark mass ranging from $\MtopMC=169~\GeV$ to  $\MtopMC=176~\gev$  are used for analysis validation.
A MC sample based on the $bb4l$ \POWHEG package~\cite{Jezo:2016ujg} %
interfaced to \PYTHIAV{8}, currently under development and scrutiny within ATLAS, is used to validate the $2\rightarrow7$ result obtained with the MC ensemble of $\ttbar$ and \singletW\ production.

The $\Zboson\text{+jets}$ production with $\Zboson \rightarrow \ee / \mumu / \tautau$  is modeled using \SHERPAV{2.2.1}~\cite{Gleisberg:2009}.
The matrix elements are calculated for up to two additional partons at NLO accuracy in QCD and four partons at LO using the \COMIX~\cite{Gleisberg:2008} and \OPENLOOPS~\cite{Cascioli:2012} matrix element generators, merged with the \SHERPA\ parton shower using the ME+PS@NLO prescription~\cite{Hoeche:2013}.
The NNPDF3.0 NNLO PDF set is used together with a dedicated parton shower tuning developed by the \SHERPA\ authors.
The samples are normalized to the total $\Zboson\text{+jets}$ cross-section at NNLO accuracy~\cite{Anastasiou:2003ds}, obtained with the \textsc{FEWZ} program~\cite{Gavin:2010az}, and a 50\% normalization uncertainty is associated to them.

The production of diboson ($\Wboson\Wboson/\Wboson\Zboson/\Zboson\Zboson$) events in association with jets is simulated using \SHERPAV{2.2.2}  with the dedicated UE tune from \SHERPA and the NNPDF3.0 \NNLO\ PDF set.
In the simulation, matrix elements at NLO accuracy in QCD for up to one additional parton and at LO accuracy for up to three additional parton
emissions are employed.
A 50\% normalization uncertainty is assigned to the diboson MC samples.

The processes $\ttbar\Zboson$, $\ttbar\Wboson$ are simulated using \MGMCatNLO\ interfaced with \PYTHIAV{8}.
The associated Higgs boson and \ttbar\ production ($\ttbar H$)  is simulated using \POWPYTHIA~8~\cite{Hartanto:2015uka}.
For the $\ttbar\Zboson$, $\ttbar\Wboson$, and $\ttbar H$ production processes, the NNPDF 3.0 NLO PDF set is used in the ME, while the A14 UE tune is applied with NNPDF2.3 LO in the parton shower.
The production processes are normalized to the corresponding NLO QCD+EW theoretical cross-sections~\cite{Florian:2016}.

All MC samples are overlaid with additional $pp$ interactions (pileup), generated with \PYTHIAV{8.186}~\cite{SJOSTRAND2008852} using the
NNPDF2.3 LO PDF set and the A3 set of tuned parameters~\cite{ATL-PHYS-PUB-2016-017}, and the average number of interactions per bunch crossing is reweighted to match that in data.
All simulated samples other than those generated with \SHERPA use the \texttt{EvtGen}~\cite{Lange:2001uf} program to simulate the decays of bottom and charm hadrons.

The number of events where at least one of the reconstructed lepton candidates originates from the decay of a hadron or a jet incorrectly identified as a lepton (fake leptons) is estimated by using the truth information associated to the reconstructed lepton in MC simulations. If the reconstructed lepton is found not to originate from a true lepton, the event is marked as a fake-lepton background event.
A \ttbar~sample with the same setting as the nominal one, but also including semileptonic top-quark decays, and the \singletW\ MC sample are used for the fake leptons estimation.
The fake leptons contribution amounts approximately to 1\% of the total event yield in the dilepton selection and a normalization uncertainty of 50\% is associated to it. The final estimate is found to be compatible within uncertainties with other data-driven estimates~\cite{TOPQ-2018-26}.

A summary of the MC generators used to simulate the most important processes can be found in Table~\ref{DataMC:tab:xSec}.
\begin{table}[!hb]
\centering
\caption{A summary of basic MC generator settings used to simulate various SM processes and the cross-section uncertainty used in their normalization.
}
\begin{small}
\resizebox{1.\textwidth}{!}{
\begin{tabular}{lcccccc}
\hline
Sample & Generator & ME PDF & Shower & Normalization & Cross-section [pb] & Norm. unc. [\%]\\
\hline
\hline
\ttbar & \POWHEG & NNPDF3.0 & \PYTHIA~8 & NNLO+NNLL & 831.8~\cite{Barnreuther:2012wtj,Czakon:2012zr,Czakon:2012pz,Czakon:2013goa} & 6.1\\
\hline
Single-top (\singletW) & \POWHEG & NNPDF3.0 & \PYTHIA~8 & (Approx)NNLO & 71.7~\cite{ATLAS-CONF-2015-039,Kidonakis:2021vob}  & 5.3\\
\hline
$Z$+jets & \SHERPA~2.2.1 & NNPDF3.0 & \SHERPA~2.2.1 & NNLO & 2107.0~\cite{Anastasiou:2003ds,Gavin:2010az} & 50\\
\hline
Diboson & \SHERPA~2.2.2 & NNPDF3.0 & \SHERPA~2.2.1 & NLO & 176.0~\cite{Campbell:1999}& 50\\
\hline
$t\bar{t}Z$ & MG5\_aMC@NLO & NNPDF3.0 & \PYTHIA~8 & NLO(QCD+EW) & 0.88~\cite{Florian:2016} & 14\\
\hline
$t\bar{t}W$ & MG5\_aMC@NLO & NNPDF3.0 & \PYTHIA~8 & NLO(QCD+EW) & 0.60~\cite{Florian:2016} & 13\\
\hline
$t\bar{t}H$ & MG5\_aMC@NLO & NNPDF3.0 & \PYTHIA~8 & NLO(QCD+EW) & 0.51~\cite{Florian:2016} & 13\\
\hline
\end{tabular}
} %
\label{DataMC:tab:xSec}
\end{small}
\end{table}

% End of text imported from the .//content/Analysis-DataMC-Signal.tex input file

%
%
%
%

%

\section{Object reconstruction and event selection}
\label{sec:ObjectsAndEventSelection}

% The next lines are included from the .//content/ObjectsAndEventSelection.tex input file

Electron candidates are identified by matching energy deposits in the electromagnetic calorimeter with a corresponding track in the inner tracking detector. Electrons are required to satisfy the \textsc{Tight} identification criteria~\cite{EGAM-2018-01} and to lie within $|\eta| < 2.47$, with the calorimeter transition region of $1.37 < |\eta| < 1.52$  being excluded.
Electron candidates
must pass the \textsc{Tight} isolation requirement~\cite{EGAM-2018-01} that removes fake and non-prompt electrons from heavy flavor decays.
Muon candidates are reconstructed by combining a track in the inner tracking detector with either a track or hits in the muon spectrometer.
Muons fulfilling \textsc{Medium} track requirements~\cite{MUON-2018-03} with $|\eta| < 2.5$  and satisfying the \textsc{Tight} isolation requirement of Ref.~\cite{MUON-2018-03} are selected.
The performance of lepton reconstruction and identification efficiency is measured in data and used to calibrate the performance in the simulations~\cite{EGAM-2018-01,MUON-2018-03,EGAM-2021-01,MUON-2022-01}.

Both electrons and muons are required to have $\pT>20~\gev$.
The transverse impact parameter divided by its estimated uncertainty is required to be less than five (three) for electron (muon) candidates and the longitudinal impact parameter must be smaller than $0.5$ mm for both lepton flavors.

Jets are reconstructed using the anti-$k_t$ algorithm~\cite{Cacciari:2008gp} with a radius parameter $R=0.4$ from particle flow objects~\cite{PERF-2015-09} that combine information from topological clusters~\cite{ATL-LARG-PUB-2008-002} and the track information for the charged component.
The jet response is calibrated with MC simulations and in situ measurements~\cite{JETM-2018-05}.
Only jets with $\pT>25~\gev$ and $|\eta| < 2.5$ are considered.
To suppress jets from pileup, the jet vertex tagger (JVT)~\cite{ATLAS-CONF-2014-018} must satisfy $\text{JVT}> 0.50$ for jets with $\pT < 60~\GeV$ and $|\eta| < 2.4$.
Jets containing $b$-hadrons are identified using the \DLOner\ algorithm~\cite{FTAG-2019-07} at the 85\% tagging efficiency working point.
Measurements of the efficiency of the $b$-tagging algorithm in data are used to adjust the simulation to match the performance measured in data~\cite{FTAG-2018-01,FTAG-2019-02,FTAG-2020-08}.

An overlap removal procedure is applied to avoid double-counting the same reconstructed objects.
Jets within $\Delta R = 0.2$ of an electron or muon are discarded and electrons within  $0.2 < \Delta R < 0.4$ of the remaining jets are rejected.  Jets with less than three tracks are removed if they overlap with a muon within $\Delta R = 0.4$.

The missing transverse momentum (\MET) is computed as the negative vectorial sum of the transverse momenta of all identified and calibrated objects in the detector, including a track-based term to account for the energy from particles not associated with a reconstructed lepton or jet~\cite{ATLAS-CONF-2018-023,PERF-2016-07}.

\subsection{Event selection}

Events are required to have at least one primary vertex with at least two associated tracks with $\pT^\text{track}>0.5~\GeV$.
The selection further requires the presence of exactly one electron and one muon of opposite electric charge, separated by $\Delta R>0.4$ . The lepton with the highest \pT is required to have $\pT>28~\GeV$ to ensure high efficiency of the single-lepton triggers.

A requirement of  $\MET> 30~\GeV$ is imposed, due to the presence of undetected neutrinos.
Exactly two $b$-tagged jets with $\pT > 30~\GeV$ are required.
Among the remaining jets, the \pT-leading one is required to have $\pT>50~\GeV$.
The relatively high \pT requirement on this jet, referred to as the \emph{extra jet} in the following, reduces the uncertainties from the jet energy calibration, the uncertainty due to the choice of parton-shower model, and the probability that the additional jet used in the \ttbarjet~system reconstruction originates from pileup.
It also reduces the theoretical uncertainties in the FO theory calculations.

Each lepton is associated a $b$-jet such that the sum of the invariant masses of the two possible lepton and $b$-jet systems ($m_{(\ell b)_1}$ + $m_{(\ell b)_2}$) is minimized. A requirement of $m_{(\ell b)_{i}} < 200~\gev$, with $i\in\{1,2\}$, is imposed on each of the two lepton and $b$-jet pairs, to avoid a region of phase space which is known to be difficult to model~\cite{TOPQ-2017-05}.

% End of text imported from the .//content/ObjectsAndEventSelection.tex input file

%
\subsection{Reconstruction of the \texorpdfstring{$t\bar t$}{ttbar} system}
\label{sec:Analysis:sub:TopReconstruction}

% The next lines are included from the .//content/Analysis-TopReconstruction.tex input file

The reconstruction of the \ttbar~system requires an estimate of the kinematic properties of the two neutrinos from the $W$ boson decays.
The neutrino four-vectors are derived from the observed \MET\ and the properties of the other objects in the event.
In this measurement, a combination of the loose kinematic reconstruction~\cite{CMS-TOP-18-004} and $\phi$-weighting
reconstruction~\cite{CDF:2009vjm} methods is employed. No top-quark mass assumption is introduced in these methods, and the combination of the two methods has a high reconstruction efficiency of around 98\%.

The loose kinematic reconstruction does not attempt to reconstruct the neutrino and anti-neutrino, but determines the kinematic properties of the $\nu\bar{\nu}$ system. The transverse momentum of the $\nu \bar{\nu}$ system is obtained directly from the \MET. %
The di-neutrino longitudinal momentum and energy are set equal to the longitudinal momentum and
energy of the charged-lepton pair.
The \ttbar\ system is then reconstructed from the two charged leptons, the two selected $b$-tagged jets, and the neutrino system in the selected events.

Around 25\% of the events reconstructed with the loose method have the reconstructed invariant mass of the \Wboson\ boson pair and neutrino pair not satisfying $m(W^+W^-) \geq 2m_W^\text{PDG}$ and $m(\nu\bar{\nu}) \geq 0$, and are hence not considered as physical solutions.
For those events, the \ttbar~system is reconstructed with an alternative approach based on the $\phi$-weighting method~\cite{CDF:2009vjm}.
For each event, each of the neutrino azimuthal angle is set to a random value, hence fixing the \pT of each individual neutrino. Then, neutrino longitudinal momenta are inferred assuming that each neutrino four-momentum and its pairing with a lepton reconstructs a $W$ boson with mass $m_W^\text{PDG}$. A $b$-jet is assigned to each lepton--neutrino pair
to compute the quantity $\chi_{\phi}^2=( m_{\nu\bar \ell b} - m_{\bar{\nu}\ell \bar{b}})/ (m_{\nu\bar \ell b} + m_{\bar{\nu}\ell \bar{b}})$, where $m_{\nu\bar \ell b}$ and $m_{\bar{\nu}\ell \bar{b}}$ are the masses of the two reconstructed top-quark candidates in the event, each obtained summing the four-momenta of one neutrino, one charged lepton and one $b$-jet. The azimuthal angle phase space $(\phi_1,\phi_2)$ of the neutrinos is scanned with 100 random values from a uniform distribution and all the combinations of neutrinos, charged leptons and $b$-jets are used in the calculation of $\chi_{\phi}^2$, hence providing 1600 possible values for it\footnote{The number of possible neutrino--lepton pairings is 8 due to the two-fold ambiguity in longitudinal momentum for each neutrino and the two lepton-neutrino combinations. An additional factor of two in the possible solutions comes from the pairing with jets.}.
The solution that minimizes the difference between the reconstructed mass of top and anti-top quark, as quantified by $\chi_{\phi}^2$, is taken for the \ttbar\ system definition.

To form the \ttbarjet\ system the extra jet is selected as the highest in \pT among the untagged jets with $|\eta| < 2.5$, and it is added to the \ttbar\ system.
The  \pT requirement on the extra jet is raised to 60~\GeV for the $2\to 7$ measurement, to match the value chosen for the FO calculation.

Figure~\ref{fig:migration_resolution}
illustrates the resolution of \Rhos achieved by the reconstruction algorithm, estimated using the nominal \ttbar{} MC simulation based on \POWPYTHIA~8, where the resolution is taken as the root mean square of the $ (\Rhos^\text{reco}-\Rhos^\text{true})/\Rhos^\text{true}$ distribution and \Rhos~is defined in Eq.~(\ref{eq:R_definition}).
The resolution of the combined loose kinematic and $\phi$-weighting reconstruction is compared with that of the individual loose kinematic and $\phi$-weighting methods alone, and to the resolution obtained with perfect neutrino reconstruction, based on MC truth information.
The gain in statistical power from combining the loose and $\phi$ methods outweighs the slight reduction in resolution.
Although the presence of undetected neutrinos affects the resolution of the reconstructed observable, the employed \ttbar~system reconstruction technique ensures the \Rhos distribution can be measured precisely, maintaining the observable resolution within the 0.05--0.1 range.

\begin{figure}[ht]
\centering
\includegraphics[width=0.6\textwidth]{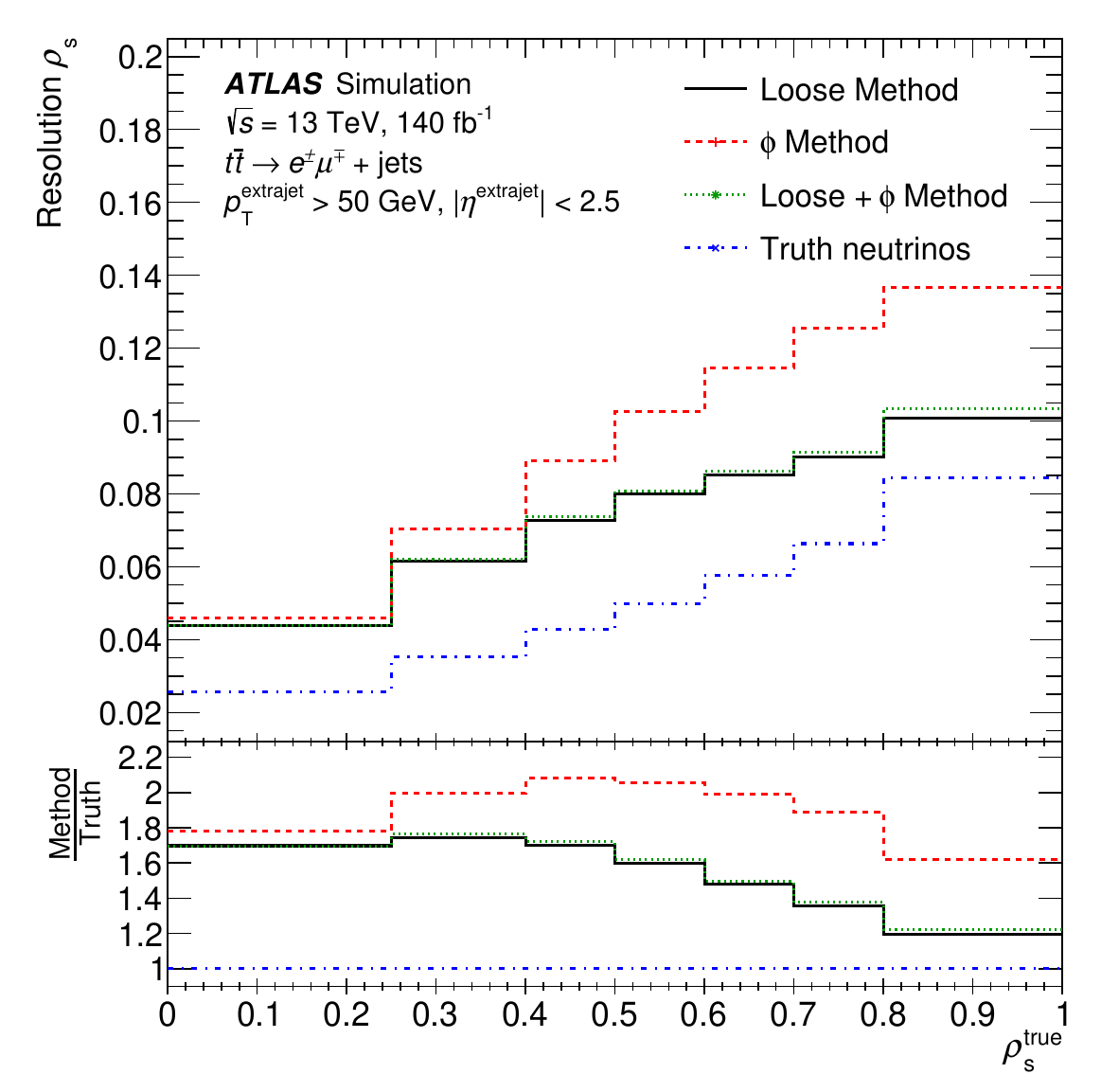}
\caption{Resolution of the $\Rhos$ observable for events reconstructed with the combined loose kinematic and $\phi$-weighting reconstruction methods, %
in the \emu decay channel of the \ttbarjet\ system, simulated using \POWPYTHIA~8.
The resolution obtained from each individual reconstruction method is also shown %
together with the best possible scenario, where the neutrino momenta are known from truth information in the MC (blue). Only events satisfying the respective reconstruction method requirements are considered.}
\label{fig:migration_resolution}
\end{figure}

The event yields in the simulation and data after all selection requirements are given in Table~\ref{table:dataMC_Rsel}.
The $t\bar{t}$ signal purity is greater than 95\%. The largest background contributions originate from single-top-quark production, fake leptons from \ttbar~and single-top-quark events and associated production of \ttbar~with a heavy boson.
Smaller contributions are expected from diboson and $Z+$jets processes.
The observed number of events agrees with the expected yield within 4\%, a difference which is covered by the uncertainty in the SM prediction, as reported in Table~\ref{DataMC:tab:xSec}.

% The next lines are included from the .//content/tables/Yields_Selection.tex input file
\begin{table}[th]
\centering
\medskip
\caption{\normalsize Event yields after the final selection for the values of minimum $\pt^\text{extrajet}$ considered in the analysis.
The uncertainties only contain the cross-section normalization components, while for the \ttbar\ sample also statistical and systematic uncertainties, as defined in Section~\ref{sec:Results:sub:SystematicUncertainties:subsub:Backgroundmodelling}, are considered.}
\resizebox{8cm}{!}{

\begin{tabular}{ccc}
\hline
& $\pt^\text{extrajet}>50~\GeV$&  $\pt^\text{extrajet}>60~\GeV$\\
\hline
\hline
$t\bar{t}$  & 103000 $\pm$ 7000 & \phantom{0}86000 $\pm$ 6000 \\
Single top  & \phantom{00}1840 $\pm$ \phantom{0}100 & \phantom{00}1470 $\pm$ \phantom{00}80  \\
Diboson     & \phantom{0000}53 $\pm$ \phantom{00}27& \phantom{0000}48 $\pm$ \phantom{00}23  \\
$Z$+jets    & \phantom{0000}55 $\pm$ \phantom{00}28& \phantom{0000}48 $\pm$ \phantom{00}24  \\
$t\bar{t}V+t\bar{t}H$   & \phantom{000}590 $\pm$ \phantom{00}80   & \phantom{000}530 $\pm$ \phantom{00}70 \\
Fake leptons & \phantom{000}750 $\pm$ \phantom{0}380&  \phantom{000}640 $\pm$ \phantom{0}320  \\
\hline
Total MC    & 106000 $\pm$ 7000 & \phantom{0}89000 $\pm$ 6000\\
\hline
Data        & 102215 & 85366  \\
\hline
\hline
Data/MC     & 0.96  & 0.96 \\
\hline
\end{tabular}

}

\label{table:dataMC_Rsel}
\end{table}

The observed $\Rhos$ distributions for the full Run~2 data set are shown in Figure~\ref{fig:R_obs_detector_bkg}.
Overall, reasonable agreement is observed between the reconstructed distribution in the data and the SM expectation generated with MC simulations of the signal and background processes.

\begin{figure}[ht]
\centering
\subfloat[\centering]{\includegraphics[width=0.44\textwidth]{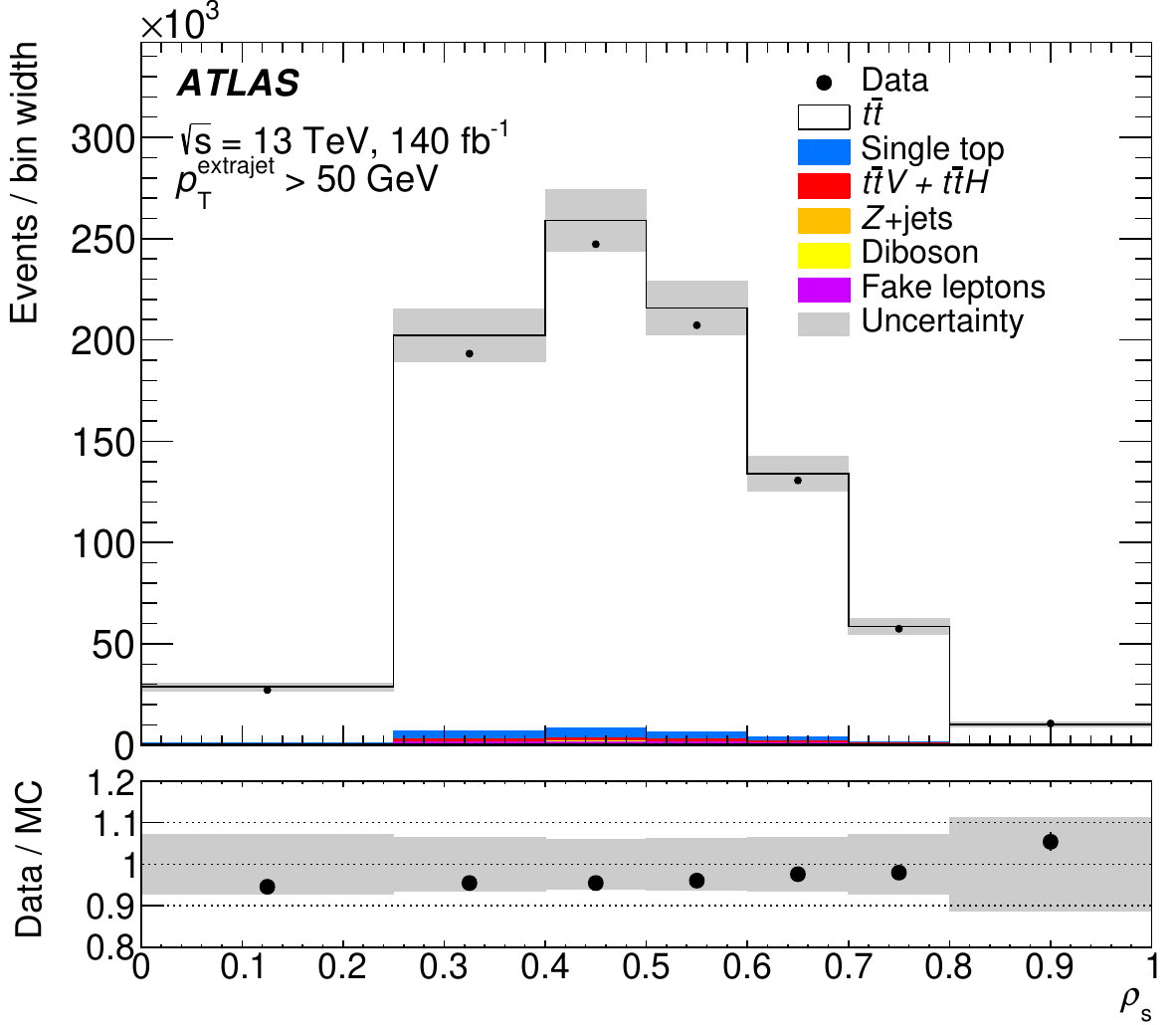}}
\subfloat[\centering]{ \includegraphics[width=0.44\textwidth]{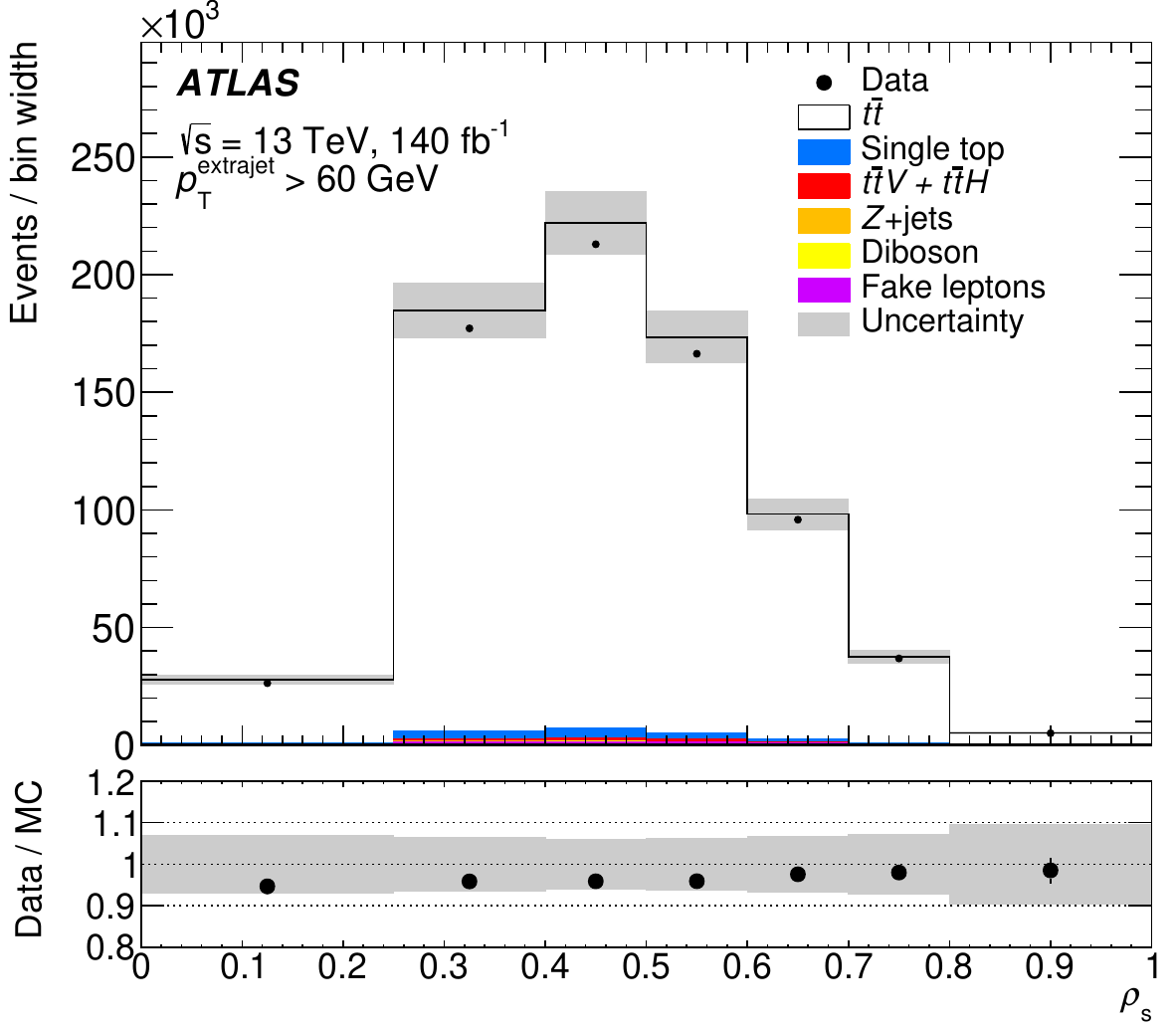}
}
\caption{Detector level distributions of the $\Rhos$ variable in the $\ttbarjet$ system after the final selection in the $\emu$ decay channel, for a cut on the extra jet \pT corresponding to (a) 50~\GeV and (b) 60~\GeV . Data (filled markers) are compared with the SM expectation with $m_t^\text{MC} = 172.5~\GeV$ (histogram). The uncertainty band represents the total uncertainty on the MC prediction, including all statistical and systematic components.
}
\label{fig:R_obs_detector_bkg}
\end{figure}

\FloatBarrier

% End of text imported from the .//content/Analysis-TopReconstruction.tex input file

\section{Parton-level definitions and fixed-order calculations}
\label{sec:PartonLevelAndFixedOrder}

% The next lines are included from the .//content/Analysis-FixedOrderCalculation.tex input file
To compare the measured observable to FO theoretical predictions, the data is corrected to  parton level.
Two FO calculations are considered. The first FO prediction for \Robs\ is computed at NLO accuracy in the strong coupling for stable top quarks. Results are obtained using the \textsc{ttbarj} process in \POWHEGBOXV{2}~\cite{Alioli:2011as}, with the on-shell top quarks in the pole-mass scheme. This calculation is referred to as $2\rightarrow3$ in the following, as three objects are considered in the final state.
The extra jet is defined from gluons and non-top quarks with the \antikt\ algorithm~\cite{Cacciari:2008gp} implemented in FastJet~\cite{Cacciari:2011ma}, with radius parameter $R=0.4$, and is required to have $\pT>50~\GeV$ and $|\eta|<2.5$.
In a previous mass measurement in \ttbarjet\ events~\cite{TOPQ-2019-12},  a similar calculation was used where the renormalization and factorization scales where fixed to the top-quark mass.
In the current measurement, dynamic scales are used in the calculation, %
which was observed to lead to reduced changes in the observable when scales were varied by factors of 0.5 and 2.
The factorization and renormalization scales are set to $\muF,\muR = \mu_0 = E_\text{T}/2$ according to the recommendations in Ref.~\cite{Alioli:2022lqo}, with $
E_\text{T} = \sum_{i\in[t,\bar{t},j]} \sqrt{{\pT^i}^2+m_i^2}$,
where $m_i$ is the invariant mass of the corresponding object.

The second prediction is a FO NLO QCD of the  $pp \rightarrow b\bar{b} l^+ \nu l^- \bar{\nu} + j$ process and is referred to as $2\rightarrow7$ in the following. This calculation includes contributions from single-resonant
and non-resonant diagrams, as well as the doubly resonant top-quark pair production process hence considering top-quark decay and full off-shell effects. The predictions for \Robs~in the $2\to 7$ approach were provided by the authors of Ref.~\cite{Bevilacqua:2016jfk}.
Only electrons and muons are considered as leptons.
Asymmetric \pT requirements are imposed on the \pT-leading and \pT-subleading leptons, with both selections matching the detector level selection.
All the objects in the seven-particle final state, excluding neutrinos,  are separated by $\Delta R>0.4$ and satisfy $|\eta|<2.5$.   A cut on the sum of neutrinos transverse momenta is set to be $\left(\sum_\nu p^{\nu}\right)_\text{T}>30~\gev$.
Jets containing a $b$-quark parton are required to satisfy $\pT^{b}>30~\GeV$, while the additional jet must satisfy $\pT^{j}>60~\GeV$.
The renormalization and factorization scale are set to $\muR,\muF = \mu_0 = H_\text{T}/2 = \frac{1}{2}\sum_{i\in [\ell^+,\ell^-, b,\bar{b}, \nu,\bar{\nu}, j]} \pT^{i}$ as suggested  in Ref.~\cite{Bevilacqua:2016jfk}.

The nominal calculations are performed with the PDF4LHC21 PDF set~\cite{PDF4LHCWorkingGroup:2022cjn}.
For comparisons of PDF sets, the calculations are also repeated with the CT18 NLO PDF set with $\alphas = 0.118$ ~\cite{Hou:2019efy},
NNPDF3.0 NLO~\cite{Ball:2014uwa}, %
ABMP16~\cite{Alekhin:2017kpj},
and the MSHT20 NLO  PDF sets~\cite{Bailey:2020ooq}. %
All considered PDF sets describe the proton content with massless $b$-quarks (five-flavor scheme).
The impact of missing higher-order corrections in the calculations is estimated with scale variations of $\muR, \muF = \{\frac{1}{2}\cdot \mu_0,2\cdot \mu_0\}$, avoiding combinations where the ratios of $\muR/\muF=\{1/4,4\}$.
The running of the strong coupling constant \alphas\ is calculated in the five-flavor scheme.
These variations are assessed for $\Mpole=172.5~\GeV$. Additional theory predictions are calculated for values of \Mpole~ranging from $169~\GeV$ to $176~\GeV$ and are interpolated with second order polynomials to produce  continuous predictions for $\Robs$.%

To correct the data to parton level, equivalent definitions are defined in the MC for the $2\to3$ and $2\to7$ calculations.
In the \ttbar MC simulation %
the $2\rightarrow3$ parton level is defined as follows: the last stable top (anti-)quark after initial state (ISR) and final state (FSR) radiation, and before decay is identified with the on-shell stable top (anti-) quark.
The additional parton-level jet is built from all partons after the parton shower and before the hadronization stage which do not originate from top (anti-)quark decays. The \antikt\ algorithm with radius parameter $R=0.4$ is used for the jet clustering.
This parton-level definition is consistently applied to both the \PYTHIAV{8} and the \HERWIGV{7} parton showers considered in the measurement.
A \ttbarjet\ event is selected for the $2\rightarrow3$ parton-level phase space if at least one additional resolved parton-level jet in $|\eta|<2.5$ and with $\pT > 50~\gev$ is found that does not originate from the decay products of the top (anti-) quarks.

For the definition at parton level in the $2\rightarrow7$ approach an ensemble of MC samples is used, as \ttbar\ production, single-top production, and di-boson production processes all contribute to the  $pp \rightarrow b\bar{b} l^+ \nu_l l^- \bar{\nu}_l + j$ process. Only electrons and muons coming from a $W$-boson decay are considered as leptons in the final state. The jets are clustered with the \antikt~algorithm on the partons after the parton shower simulation and before the hadronization step.
The $b$-jets are identified by requiring that the direction of the momentum of the $b$-quark, from which the jet stems from, lies within $\Delta R=0.4$ of the jet.
Selection cuts on the objects are imposed such that kinematic requirements are synchronized across the FO theory prediction~\cite{Bevilacqua:2016jfk}, the MC parton-level definition and the event selection at the detector level.
The  $\rho_s$ variable in the $2\rightarrow7$ parton-level is obtained by adding the four-vectors corresponding to the charged leptons and neutrinos from the $W$ boson decays, the two \pT-leading $b$-jets and the \pT-leading additional jet.

%
%
%
%
%
%

% End of text imported from the .//content/Analysis-FixedOrderCalculation.tex input file

%

\section{Differential cross-section measurement}
\label{sec:DifferentialCrossSection}
%
%
%
%

% The next lines are included from the .//content/Analysis-Unfolding.tex input file
The differential cross-section \diffXStext\ is obtained by subtracting the background contributions and unfolding to the $2 \to 3$ and $2 \to 7$ parton levels using the IBU algorithm~\cite{DAgostini:1994fjx}.
The background subtraction removes all the expected non-\ttbar\ processes in the $2\rightarrow3$ measurement. In the $2\rightarrow7$ case instead, \singletW\ production is considered as signal and is not subtracted, since \singletW\ and \ttbar\ can produce the same final state once the $W$ bosons and the top quarks decay\footnote{This is also valid for $W^{+}W^{-}$ boson production, but its contribution is negligible in this analysis.}.
The correction to parton level is split into three terms derived from the nominal \POWPYTHIA~8 simulation.
\begin{itemize}
\item An acceptance correction considers the events that satisfy the detector-level selection, but do not fulfill the parton-level definition.
This correction is implemented as a bin-by-bin factor, defined as the ratio of events selected at the detector level that satisfy the parton-level requirements, divided by the total number of events selected at the detector level.
\item A matrix describes the migration of events that satisfy both detector and parton level selections, from bins at the detector level to the bins at the parton level.
\item An efficiency correction %
accounts for parton-level events that fail to meet the detector-level selection.
This correction is also implemented as a bin-by-bin factor, computed as the ratio of parton level events which fulfill both parton and detector level selections over parton level events which only need to satisfy the parton level requirements.

\end{itemize}
The binning of the differential cross-section is optimized such that the high-$\rho_s$ region, that has the highest sensitivity to \Mpole\, is isolated. At the same time, this region should have a large enough population to ensure less than 5\% statistical uncertainty in the high-sensitivity bins of the detector-level distribution. Finally, migrations from bins at detector level to parton level are limited, keeping the diagonal elements of the migration matrix above 40\%.
The migration matrix, the efficiency, and the acceptance for the $2\rightarrow3$ parton level unfolding are shown in Figure~\ref{fig:migration_effacc}.
The low values of the efficiency term reflects the fact that the parton level definition extrapolates for non-dileptonic top-quark pair decays and covers a much more inclusive phase space than the detector level one.

\begin{figure}[ht]
\centering
\subfloat[\centering]{\includegraphics[width=0.5\textwidth]{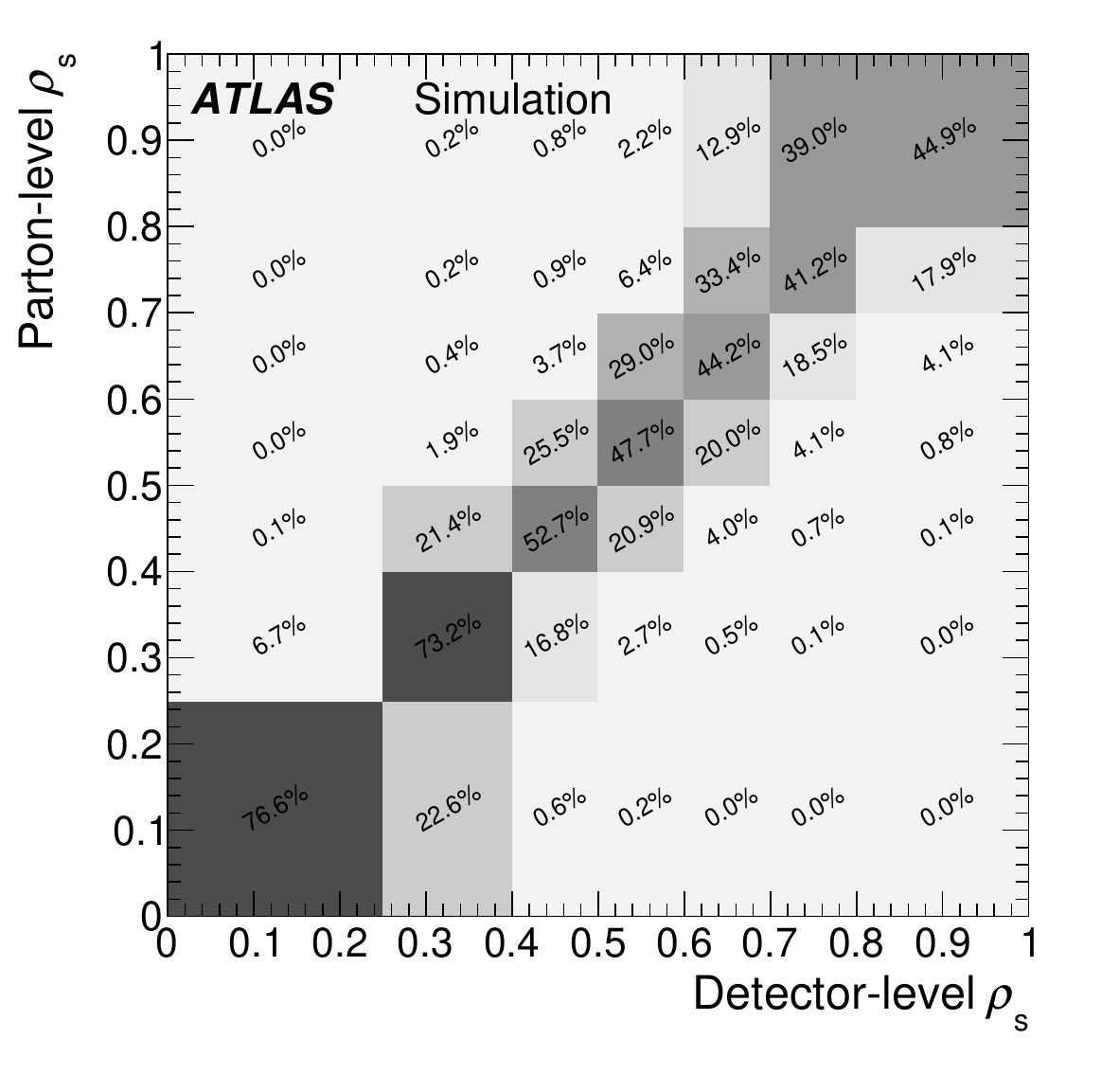}}

\subfloat[\centering]{\includegraphics[width=0.45\textwidth]{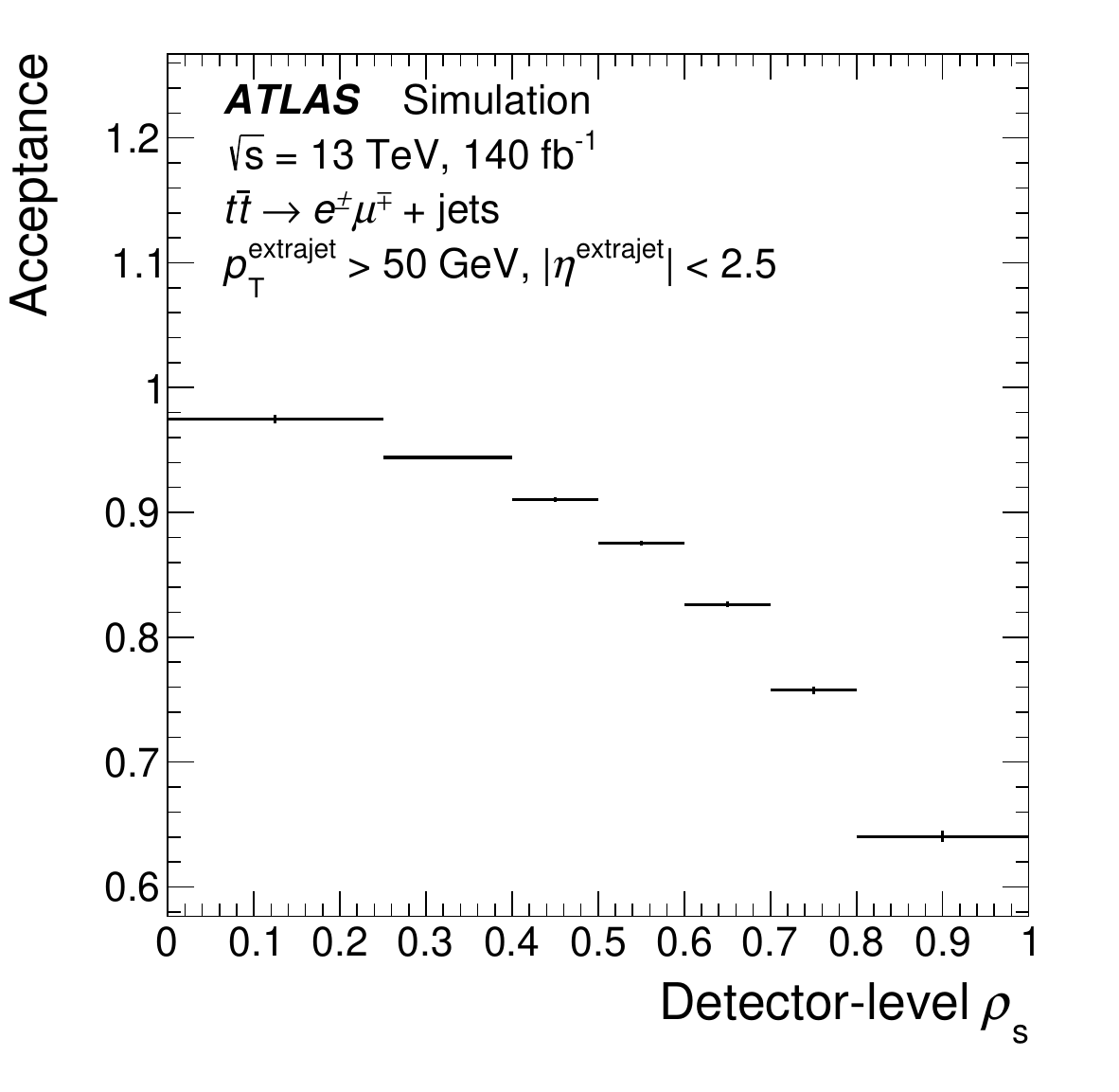} }
\subfloat[\centering]{\includegraphics[width=0.45\textwidth]{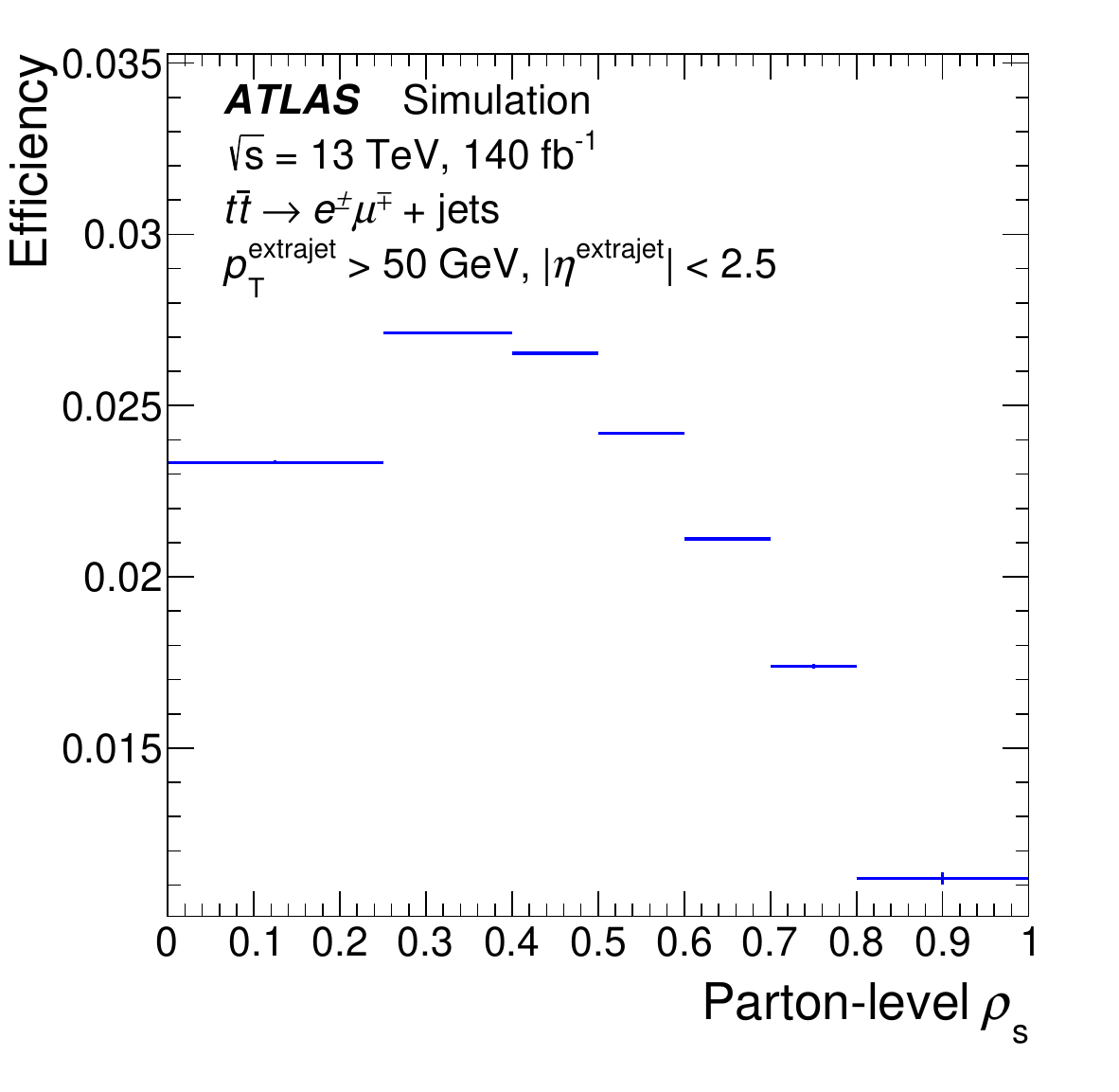} }
\caption{The (a) migration matrix and (b) the acceptance and (c) efficiency factors for events considered in the unfolding to the $2\rightarrow3$ parton level.
The matrix and correction factors are built from the nominal simulation of \ttbarjet\ events using the \POWPYTHIA~8 generator. Vertical bars represent MC statistical uncertainties.}
\label{fig:migration_effacc}
\end{figure}

The IBU procedure starts from the parton-level distribution of the nominal signal simulation. It iteratively applies a correction to the parton-level based on the unfolding matrix and the measured spectrum at detector level. These iterations progressively reduce the bias.
The number of iterations, and hence the level of regularization, is chosen such that the unfolding procedure does not depend on the value of the top-quark mass. To test this, MC samples with the top-quark mass parameter set to values between $\MtopMC=169$~\gev\ and  $\MtopMC=176~\gev$ are unfolded.
Good convergence with minimal bias is found for all test distributions with 70 (40) iterations for the correction to the $2\rightarrow3$ ($2\rightarrow7$) parton level. These settings are used for all the results presented in the following.

The covariance matrix $V^\text{stat}$ encodes the statistical uncertainty associated with the unfolded differential cross-section. It is built from 50000 pseudo-data distributions generated by varying background-subtracted \diffXStext\ distribution within its statistical uncertainty, which are unfolded with the nominal unfolding procedure.
Experimental systematic effects are added to the statistical covariance matrix to build a full covariance matrix, as explained in detail in Section~\ref{sec:SystematicUncertainties}.

%
\begin{comment}
The statistical uncertainty on the unfolded differential cross-section is specified with a covariance matrix $V_{ij}$ that considers the correlations between the bins in the unfolded distribution.
The statistical uncertainties on \diffXStext\ and the uncertainty due to limited MC statistics in the migration matrix $M$ are added to construct the covariance matrix $V_{ij}$
\begin{equation} \label{eq:covmat}
V_{ij} = V_{ij}^{\text{data}}( \RobsDet_{\text{data}}(\Rhos) ) + V_{ij}^{\text{MC}}(M).
\end{equation}
\end{comment}
The unfolded \diffXStext\ distribution is normalized to unity to obtain the \Robs\ observable at parton level which is used to extract the top-quark pole mass.
The covariance matrix of the normalized differential cross-section is obtained using a Cholesky decomposition~\cite{gentle1998numerical, CholeskiDecomposition:NashJ90} %
with 50000 toys, as described in Ref.~\cite{TOPQ-2019-23}.

The results for the normalized differential cross-sections at parton level are presented in Figure~\ref{fig:result_diffXS}. In the unfolding to the $2 \to 3$ parton-level, the measurement reaches a precision of 8\% in the $\rho_s$ bin from 0.7 to 0.8 and 30\% in the last bin from 0.8 to 1. The $2 \to 7$ phase space definition is closer to the detector level one and the measurement reaches a slightly better precision of 7\% and 24\% in the next-to-last and last bins. The predictions for two values of the top-quark pole mass are overlaid on the corrected data, showing the pronounced sensitivity to the top-quark pole mass in these two bins.

\begin{figure}[ht]
\centering
\subfloat[\centering]{\includegraphics[width=0.45\textwidth]{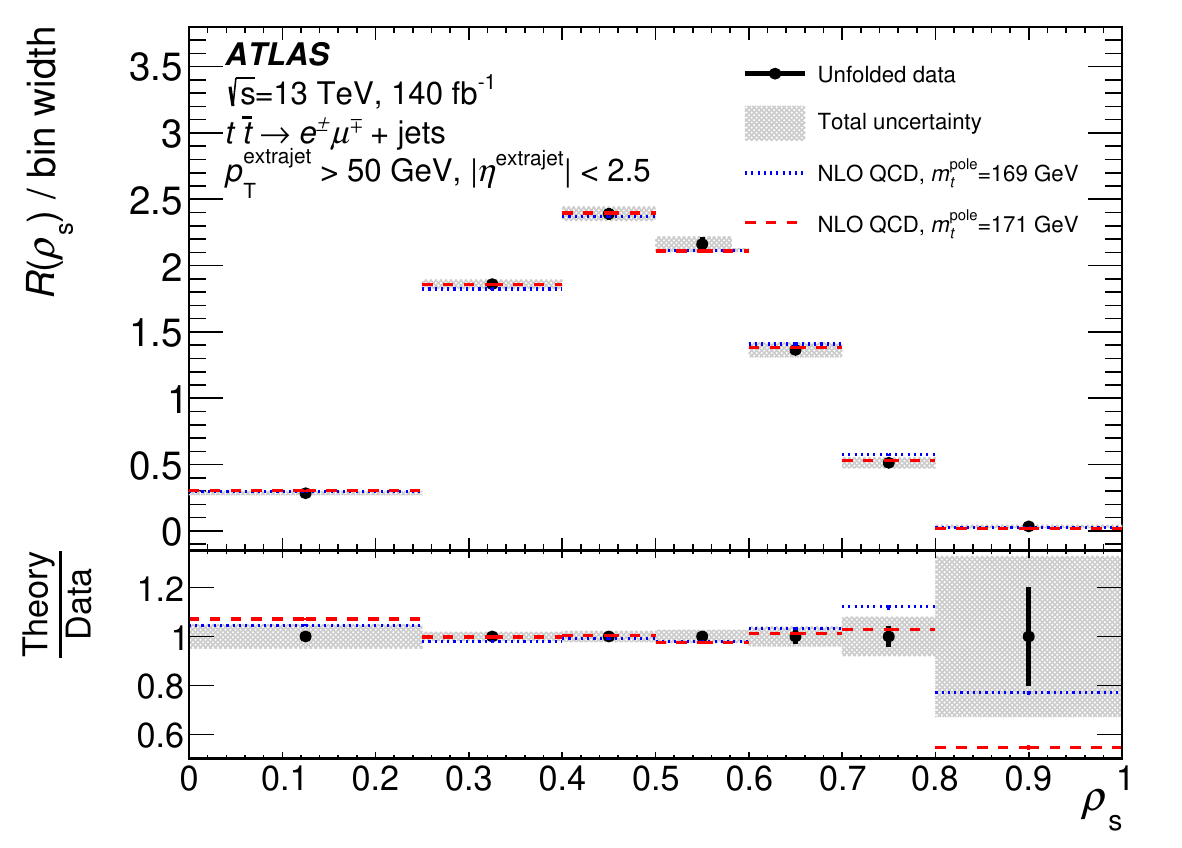}}
\subfloat[\centering]{ \includegraphics[width=0.45\textwidth]{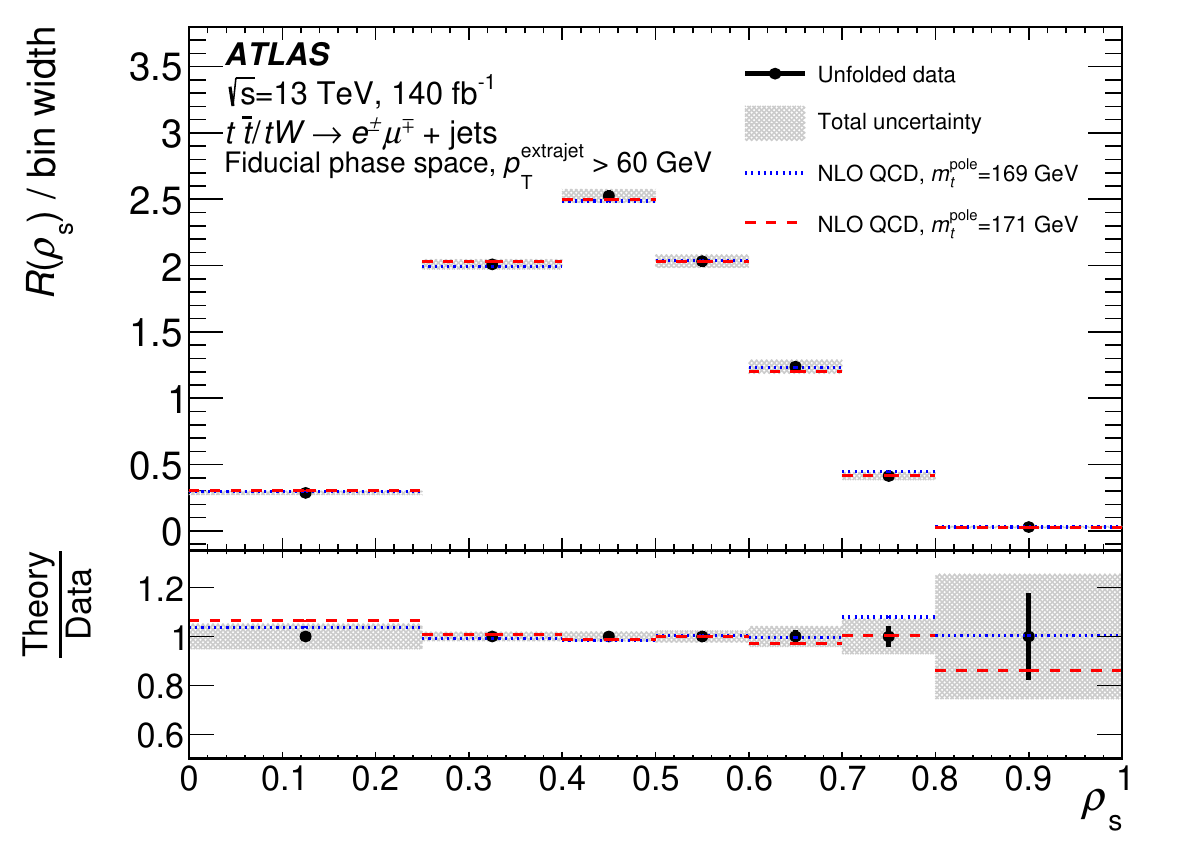} }
\caption{
The measured normalized differential cross-section \Robs\ unfolded to (a) the $2\rightarrow3$ parton level and (b) the $2\rightarrow7$ parton level.
The error bars on the marker indicate the statistical uncertainty, the gray band the total experimental uncertainty.
Theoretical predictions at fixed-order NLO QCD  for
$\Mpole=169~\gev$ (dotted line) and $\Mpole=171~\gev$ (dashed line) are also  shown, without their associated uncertainties.
}
\label{fig:result_diffXS}
\end{figure}

% End of text imported from the .//content/Analysis-Unfolding.tex input file

%

% The next lines are included from the .//content/SystematicUncertainties.tex input file

\section{Systematic uncertainties}
\label{sec:SystematicUncertainties}

% The next lines are included from the .//content/Results-SystematicUncertainties.tex input file

To evaluate the systematic uncertainties in the differential cross-section measurement, the nominal unfolding procedure is repeated on alternative pseudo-data distributions, corresponding to each systematic variation.
The impact of each systematic uncertainty in the bin $i$ of the unfolded distribution, $\Delta_{i}^\text{syst}$, is used to build a covariance matrix associated with that systematic effect, $V^\text{syst}_{i,j}=\Delta^\text{syst}_{i}\Delta^\text{syst}_{j}$.
If the uncertainty source is obtained by comparing two simulations, the full difference to the nominal result is defined as the uncertainty, and the opposite-sign variation is assumed to be of identical size.
For systematic uncertainty sources that consist of two-sided variations,
a symmetric  uncertainty is estimated as half of the difference between the up and down variations.
If both variations point in the same direction compared with the central result, the maximum is taken as the resulting uncertainty and the opposite-sign variation is assumed to be of identical size.

A separate covariance matrix is constructed for each of the systematic variations and for the statistical uncertainty. The total covariance matrix is obtained adding up all covariance matrices.
In this approach all the systematic components are independent of the other components and every individual systematic uncertainty is fully correlated across all bins of the unfolded distribution.

% End of text imported from the .//content/Results-SystematicUncertainties.tex input file

%

%
%
%

\subsection{Experimental uncertainties}
\label{sec:Results:sub:SystematicUncertainties:subsub:Detectormodelling}

% The next lines are included from the .//content/Results-SystematicUncertainties-Dec.tex input file
Uncertainties originating from the calibration of the objects used in the analysis are evaluated by unfolding the detector-level distribution obtained from MC simulations where the calibrations are varied within their uncertainties.
The signal $\ttbarjet$ and background contributions are varied simultaneously for each experimental uncertainty source.

The following uncertainty sources are considered:
\begin{itemize}
\item The uncertainty in the integrated luminosity of the Run 2 \pp\ collision data at $\sqrt{s} = 13~\TeV$ is %
0.8\%~\cite{DAPR-2021-01}.
The expected yields of all signal and background processes are varied simultaneously. The uncertainty arising from the imperfect modeling of pileup in the data is estimated by reweighting the pileup profile in all simulated signal and background processes simultaneously~\cite{STDM-2015-05}. %
\item To estimate the impact of uncertainties in the lepton reconstruction, identification and isolation efficiencies, the MC simulation is varied as a function of lepton \pt and $\eta$, in a  range allowed by measurements performed in $Z \rightarrow \ee/\mumu$ and $J/\psi \rightarrow \ee/\mumu$ events~\cite{EGAM-2021-01,MUON-2018-03}. Similarly, the
trigger efficiencies are varied to account for uncertainties in the measurements of the efficiencies of the lepton triggers~\cite{TRIG-2018-05,TRIG-2018-01}.
The effects of the uncertainties in the electron (muon) energy (momentum) scale and resolution are taken into account by varying these within their uncertainties~\cite{EGAM-2018-01,MUON-2022-01}.
\item Uncertainties in the energy scale of the jets (JES) are taken into account by varying the JES in the simulation.
The JES uncertainty is split into 35 components, originating from the in situ calibration, the pileup correction, the dependence on the jet particle content, the punch-through modeling, and the high-\pT\ jet response~\cite{ATL-PHYS-PUB-2022-021,STDM-2020-04}. The uncertainty due to the jet energy resolution (JER) is evaluated by varying the resolution within its uncertainties as a function of $\pT$ and $\eta$.
The total JER response uncertainty is split into thirteen one-sided sources which are symmetrized to estimate the effect on the measurement.
The efficiency of the JVT tagger is corrected in simulation, following a
measurement that determines the JVT efficiency with a total uncertainty of $1-2\%$ using the tag-and-probe method in $Z\rightarrow \mumu$ events with additional jets ~\cite{ATLAS-CONF-2014-018}.
\item Uncertainties originating from $b$-tagging are obtained from in situ efficiency measurements~\cite{FTAG-2018-01,FTAG-2019-02,FTAG-2020-08} %
and are propagated through the analysis.
The $b$-tagging uncertainties include different sources, split among $b$-tagging efficiency, and the $c$- and light-flavor mis-tagging rates.
\item The effects of systematic uncertainties in the jets and leptons are propagated to the \MET. Additional uncertainties for the scale and resolution of the track-based term~\cite{ATLAS-CONF-2018-023,PERF-2016-07} are also included.
\end{itemize}

% End of text imported from the .//content/Results-SystematicUncertainties-Dec.tex input file

%
\subsection{Modeling uncertainties}
\label{sec:Results:sub:SystematicUncertainties:subsub:Backgroundmodelling}

% The next lines are included from the .//content/Results-SystematicUncertainties-BackMod.tex input file
To evaluate the uncertainties in modeling of the simulated events,  the internal parameters of the nominal MC simulation are varied and alternative MC generators are employed.
To account only for changes in the response at the detector level and avoid convolution of those effects with any parton-level difference, the parton-level spectrum of the alternative MC simulations is reweighted to the one of the nominal \POWPYTHIA~8 simulation.

Many variations are considered to evaluate the uncertainty in the modeling of the \ttbar\ process.
\begin{itemize}
\item The uncertainty due to the choice of the matching procedure between matrix element and the parton shower is evaluated by varying the hardness criterion ($\pThard$) in the \POWPYTHIA~8 \ttbar\ simulation from the event scale ($\pThard=0$) to the lowest $\pT$-value between the \POWHEG\ emission, and the inital- and final-state partons ($\pThard=1$).
\item The impact of the choice of the damping parameter value in the \POWHEG\ simulation is estimated by setting $\hdamp=3\cdot \MtopMC$ and comparing the results to the ones obtained with the nominal choice of $\hdamp=1.5 \cdot\MtopMC$.
\item To evaluate the impact of missing higher orders in the QCD perturbative calculation of the matrix element, the \muR and \muF scales in the hard-process MC generator are varied  by factors of $0.5$ and $2$ in the \POWPYTHIA\ 8 \ttbar simulation, independently. %
\item The dependence of the measurements on the choice of the PS and hadronization models in the simulation is estimated by interfacing the \POWHEG\ \ttbar\ simulation with \HERWIGV{7.1.3}~\cite{Bellm:2015jjp} using the NNPDF 3.0 NLO  PDF set in the ME, and MMHT 2014 LO~\cite{Harland-Lang:2014zoa} in the parton shower and the multi-parton interaction model.

\item The uncertainty of the \alphas\ value in the multi-parton interactions and the color-reconnection of the beam remnants in \PYTHIAV{8} is propagated by varying the \textsc{Var1} set of parameters of the A14 eigentunes~\cite{ATL-PHYS-PUB-2014-021}.
\item  The uncertainty associated with the modeling of ISR is evaluated by varying the \alphas\ value in the \PYTHIAV{8} parton shower \textsc{Var3c} set of parameters, as per the A14 UE tune variations.
To estimate the FSR uncertainty, renormalization and factorization scales in the PS are varied by factors of $0.5$ and $2$.
\item The uncertainty in the modeling of the radiation in top-quark decays is evaluated by comparing the results of the \textit{recoil-to-top} scheme to one implemented from the nominal simulation, where the second and subsequent gluon emissions recoil against the $b$-quark (\textit{recoil-to-colored})~\cite{TOPQ-2017-17}.
\item The uncertainty associated with the color-reconnection modeling is estimated from two alternative models~\cite{ATL-PHYS-PUB-2017-008}, by taking as the alternative sample the one producing the largest difference on the extracted top-quark mass relative to the nominal simulation.
\item The uncertainty related to the modeling of top-quark decay is evaluated by comparing the nominal sample to a MC simulation with the same simulation settings, but where the decay of the top-quark is performed with \textsc{MadSpin}~\cite{madspin1,madspin2}. This uncertainty is referred to as \emph{line shape} uncertainty, as it impacts the shape of the top-quark mass distribution.
\item The uncertainty of the PDFs are computed using the thirty variations and the nominal version of the PDF4LHC15 PDF set at NLO accuracy~\cite{PDF4LHC2016}.
\item  An additional uncertainty is considered to cover for the effect of \ttbar NNLO corrections,  which are known to be not fully covered by the scale uncertainties of the \ttbar NLO+PS prediction~\cite{Czakon:2012zr,Czakon:2012pz,Czakon:2013goa}. An iterative reweighting procedure is implemented to simulate NNLO+NLL QCD effects in the parton-level $\pT^t$, $\pT^{\ttbar}$, and $m_{\ttbar}$ distributions.
The difference between results obtained with the reweighted and the nominal distributions is taken as an uncertainty.

\end{itemize}

Single top-quark production is the largest non-\ttbar\ process contributing to the selected events. A systematic uncertainty in the sample normalization is assigned from \singletW\ cross-section uncertainty. In addition, a dedicated systematic uncertainty is assigned to estimate the impact of the \ttbar\ and \singletW\ overlap removal: the diagram-subtraction~\cite{White:2009yt} and diagram-removal~\cite{Powheg-PhysJC71} schemes are compared with each other and the difference between results is taken as the uncertainty.

As prefaced in Section~\ref{sec:Analysis:sub:DataAndMC}, samples corresponding to other backgrounds are assigned a cross-section normalization uncertainty, indicated in Table~\ref{DataMC:tab:xSec}, while the fake-lepton background is assigned a rate uncertainty of 50\%. These uncertainties are propagated by generating pseudo-data with varied normalizations of the background processes. The nominal correction procedure is applied and the shift in the corrected differential cross-section is taken as the uncertainty.

The residual dependence of the unfolding on the top-quark mass parameter in the MC is accounted for by comparing two MC simulations generated with $\MtopMC=171~\gev$ and $\MtopMC=174~\gev$. The bin-by-bin difference between the unfolded distributions is subtracted by the expected difference from the parton level and the result is used to define an additional systematic uncertainty which covers the residual \MtopMC~dependence of the unfolding.
%

% End of text imported from the .//content/Results-SystematicUncertainties-BackMod.tex input file

%
%
%

%
%

%
%

%
%

% The next lines are included from the .//content/Results-SystematicUncertainties-MCStat.tex input file
To evaluate the uncertainty due to the limited size of the \ttbar\ MC samples used in the data correction procedure,
5000 pseudo-experiments are performed where the unfolding matrix is varied within its statistical uncertainty.
%
%
%

% End of text imported from the .//content/Results-SystematicUncertainties-MCStat.tex input file

% End of text imported from the .//content/SystematicUncertainties.tex input file

\section{Top-quark pole mass extraction}
\label{sec:PoleMassFit}

% The next lines are included from the .//content/Analysis-Fit.tex input file

To extract the value of the top-quark pole mass, a fit of \Robs\ as given by the $\ttbarjet~$ calculation at NLO QCD accuracy to the unfolded normalized differential cross-section $\RobsParton_{\text{data}}$ is performed.
The fit uses the least-squares method, where the best-fit value of $\Mpole$
is the one that minimizes:
\begin{equation}
\label{eq:chi2}
\chi^2 = \sum_{i, j} \left[\RobsParton_{\text{data}} - \RobsParton_\text{theo} (\Mpole)\right]_i \left[\Tilde{V}^{-1}\right]_{ij} \left[\RobsParton_{\text{data}} - \RobsParton_\text{theo}(\Mpole)\right]_j,
\end{equation}
where indices $i$ and $j$ refer to the bin number, and $\Tilde{V}$ is the covariance matrix of the normalized unfolded spectrum, which includes experimental and modeling uncertainties.
For the nominal fit, the PDF4LHC21 PDF set~\cite{PDF4LHCWorkingGroup:2022cjn} is used, which combines information from the CT18~\cite{Hou:2019efy}, NNPDF3.1~\cite{NNPDF:2017mvq} and MSHT20~\cite{Bailey:2020ooq}  PDF sets.
Since the $\Robs$ observable is normalized, the entries in the bins are not independent of each other and the normalized covariance matrices cannot be inverted. To calculate the $\chi^2$, the matrix is reduced by removing one column and the corresponding row, as the relative element of the residual vector. As a nominal choice, the first bin is excluded in the sum of Eq.~\ref{eq:chi2}, but the result does not depend on the choice of the excluded bin.

%
%

% The next lines are included from the .//content/Analysis-Closure_Test.tex input file
\subsection{Linearity of the unfolding}
To ensure that the measurement of the cross-section does not depend on the top-quark mass in the simulation used to correct the data, \POWPYTHIA~8 simulations are used to generate detector-level \Rhos\ distributions %
corresponding to \MtopMC\ values ranging from 169~\GeV to 176~\GeV.
Each of the mass-varied samples is then unfolded to parton level with the nominal $\MtopMC=172.5~\gev$ simulation.
A MC–based parton-level template is defined, with a continuous parameterization as a function of \MtopMC, constructed similarly to the FO theoretical template. This template is then used in a fit to the unfolded mass-varied distributions to extract a value of the top-quark mass, $m_t^\text{fit}$, for each distribution.

In Figure~\ref{fig:mass_linearity}, the results of this test are shown.
To quantify the bias introduced by the unfolding procedure, a linear fit to the values of $m_t^\text{fit}-\MtopMC$ is performed, considering only MC statistical uncertainties.
This linear fit has a slope covering a $\pm170~\MeV$ $(\pm300~\MeV)$ range  in the $2\rightarrow3$ ($2\rightarrow7$) measurement, at the extreme mass points considered. Such variations are covered by the residual \MtopMC~dependence systematic uncertainties.

\begin{figure}[ht]
\centering
\subfloat[\centering]{\includegraphics[scale=0.4]{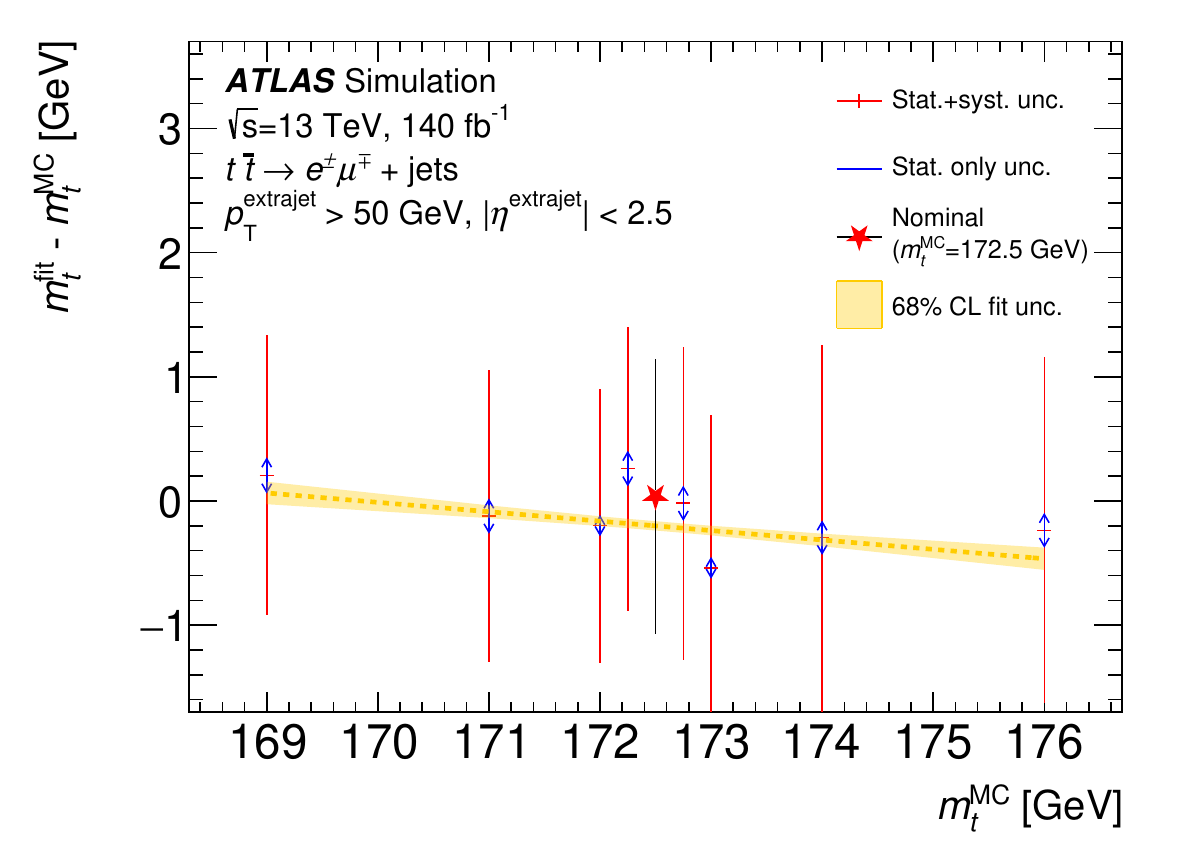}}
\subfloat[\centering]{ \includegraphics[scale=0.4]{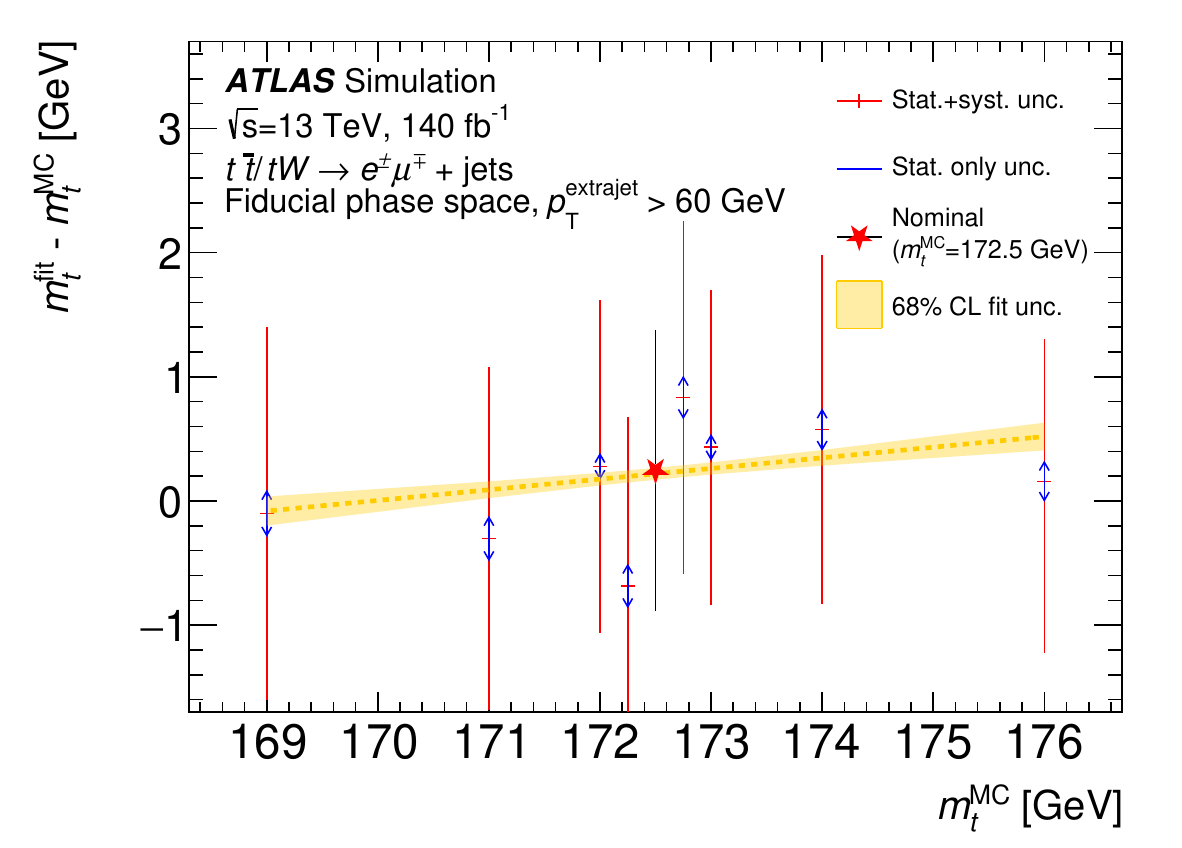}}
\caption{Difference between the extracted $m_t^\text{fit}$ value in the $\chi^2$ fit and the mass value $m_t^\text{MC}$ in the \POWPYTHIA~8 \ttbar\ simulations, (a) for the $2\rightarrow3$ and (b) $2\rightarrow7$ measurements.
Each point corresponds to a mass value extracted using the parameterized mass dependence of the parton-level distributions obtained from the simulated MC samples, with the inner uncertainty bars representing the statistical uncertainties due to the limited size of the simulated samples and the outer bars representing the total experimental uncertainty associated with that measurement.
A linear fit is performed across the various mass points, considering only their fully uncorrelated statistical uncertainties. Another uncorrelated effect, which originates from the interplay of statistical and systematic effects in the covariance matrices, further scatters the fitted mass values and is not considered in the linear fit. The 68\% confidence level (CL) associated with the fit is shown with a filled area.  The 172.5 \GeV~mass point is not included among the fitted points.}
\label{fig:mass_linearity}
\end{figure}

% End of text imported from the .//content/Analysis-Closure_Test.tex input file

% End of text imported from the .//content/Analysis-Fit.tex input file

\subsection{Theoretical uncertainties}
\label{sec:Results:sub:SystematicUncertainties:subsub:Theoreticaluncertainties}

% The next lines are included from the .//content/Results-SystematicUncertainties-Theo.tex input file
Since \Mpole\ is extracted from a $\chi^2$ fit to fixed-order theory calculations at NLO accuracy, additional uncertainties due to missing higher-order corrections and the proton PDFs and the strong coupling \alphas\ are included.
The uncertainties in the theory calculations due to the missing higher order corrections are estimated with the seven-point variation of the \muR, \muF\ scales: \muR and \muF\ in the theory calculation are varied by factors of $0.5$ and $2$, avoiding ratios of where $\muR/\muF=4$ or $\muR/\muF=1/4$. The \Mpole\ result obtained with the nominal scale choice is compared with all seven variations and the largest upward and downward variations of the mass are taken as the uncertainty.

To evaluate the uncertainty due to imperfect knowledge of the proton PDFs used in the fixed-order calculations, predictions are obtained using all the variations of the corresponding PDF set.  The PDF uncertainties are presented at the 68\% confidence level and are taken in addition to the ones affecting the unfolding and described in Section~\ref{sec:SystematicUncertainties}.
The value of the QCD coupling \alphas\ in the nominal FO calculation is chosen to be $\alphas(M_Z)=0.118$.
To estimate the uncertainty in the choice of \alphas, the value is varied between $\alphas(M_Z)=0.117$ and $\alphas(M_Z)=0.119$.
The difference to the nominal result is added in quadrature to the PDF uncertainty to form the total PDF$\oplus\alphas$ uncertainty.

%
%
%
%
%
%
%
%

% End of text imported from the .//content/Results-SystematicUncertainties-Theo.tex input file

%

%
%

% End of text imported from the .//content/Analysis.tex input file

%

\section{Results}
\label{sec:Results}

% The next lines are included from the .//content/Results-MassValue.tex input file

The top-quark pole mass is extracted from the normalized differential cross-section with a least-square fit to FO predictions at NLO accuracy, as discussed in Section~\ref{sec:PoleMassFit}.
The fit yields a $\chi^2$ per degree of freedom $\chi^2/\text{ndf}$ of 0.88, indicating good agreement between the data and the prediction. The result obtained for the top-quark pole mass in the $2\rightarrow3$ approach is:
\begin{equation}
\label{eq:massresult}
\topMassResultTotalChannel.
\end{equation}
This is the main result of this study. The first uncertainty represents the statistical component, originating from the limited size of the recorded Run 2 data sample.
Systematic uncertainties in the modeling of the detector response and signal and background processes are included in the second uncertainty contribution, which also includes the statistical uncertainty due to the finite size of the simulated samples. The third and fourth uncertainty components of Eq.~(\ref{eq:massresult}) cover the impact of missing higher-order corrections, the choice of the proton PDF set and the QCD coupling constant value in the NLO FO calculation.

A detailed breakdown of the uncertainties in the \Mpole\ measurement is presented in Table~\ref{tab:MtopTotalSystUnc_SRemu3j2b85_OnShell}.
The dominant experimental uncertainties are those related to the $b$-tagging and jet-response calibration, which amount to approximately 440~\MeV\ and 650~\MeV, respectively.
Several aspects of \ttbar\ modeling lead to significant uncertainties, in particular the difference between the recoil-to-top and recoil-to-colored schemes in the parton shower
and the parton shower generator choice.
The theoretical uncertainties from variations of the dynamical renormalization and factorization scales amount to 350--450~\MeV.
The PDF uncertainties in the calculation, evaluated using the error sets of the PDF4LHC21 combination of PDF sets, add a similar uncertainty.

\begin{table}[ht!]
\small
\centering
\medskip
\caption{Summary table of the uncertainties on the measurement of \Mpole\ from a least square fit to the $ 2 \to 3$ calculation of Ref.~\cite{Alioli:2022lqo}. The detector uncertainties grouped under \textit{Others} correspond to the sum in quadrature of missing transverse energy, pileup re-weighting, and luminosity uncertainties.
The factorization scale is responsible for the largest changes in the theoretical prediction and its high variations ($\mu_F \times 2$), yield a smaller value of the top-quark pole mass.
The $b$-tagging and jet-energy response uncertainties are made up of multiple contributions, with the leading jet-energy systematic uncertainties coming from the dependence of the jet energy response on MC modeling and pileup. The MC statistical uncertainty associated with the systematic shift is given in the third column.}
\resizebox{9cm}{!}{

% The next lines are included from the .//content/tables/Mtop_TotalSystUnc_SRllbar3j2b85_RobsOpt7Luis.tex input file

\begin{tabular}{lcc}
\hline
Uncertainty source & $\Delta m_{t}^\text{pole}$ [GeV] & MC stat. unc. [GeV]\\
\hline
\hline
Data statistics                     & 0.33  & -\\
\hline
\multicolumn{3}{c}{Detector unc.} \\
\hline
$b$-tagging and mistag              & 0.44  & 0.06\\
Jets         & 0.65  & 0.06\\
Leptons                             & 0.25  & 0.06\\
Others                              & 0.18  & 0.06\\
\hline
\multicolumn{3}{c}{Modeling unc.} \\
\hline
MC statistical uncertainty                    & 0.08  & -\\
Backgrounds normalization           & 0.02  & -\\
Single-top modeling                & 0.03  & 0.06\\
$m_{t}^\text{MC}$ dependence             & 0.10  & 0.09\\
PS Recoil model                     & 0.68  & 0.06\\
Parton shower                       & 0.43  & 0.14\\
Underlying event                    & 0.39  & 0.12\\
Color reconnection                  & 0.13  & 0.08\\
ME+PS matching: $p_\text{T}^\text{hard}$      & 0.09  & 0.06\\
ME+PS matching: $h_\text{damp}$          & 0.26  & 0.06\\
ME+PS matching: line shape          & 0.38  & 0.12\\
3D NNLO reweight                    & 0.21  & 0.06\\
PDF                                 & 0.26  & 0.06\\
Initial-state radiation             & 0.24  & 0.06\\
Final-state radiation               & 0.04  & 0.16\\
Factorization scales                & 0.09  & 0.06\\
Renormalization scales              & 0.03  & 0.06\\
\hline
\multicolumn{3}{c}{Theory unc.} \\
\hline
Scale variations                    & +0.34  -0.28 & +0.05  -0.06 \\
PDF $\oplus~\alpha_\text{S}$             & 0.24 & +0.06  -0.06 \\
\hline
\hline
Total                               & +1.47  -1.44 &  - \\
\hline
\end{tabular}

% End of text imported from the .//content/tables/Mtop_TotalSystUnc_SRllbar3j2b85_RobsOpt7Luis.tex input file

}
\label{tab:MtopTotalSystUnc_SRemu3j2b85_OnShell}
\end{table}

The value of the mass is also reported separately for several PDF sets, as recommended by the PDF4LHC group~\cite{PDF4LHCWorkingGroup:2022cjn}:
\begin{equation}
\begin{split}
& \Mpole (\text{CT18~\cite{Hou:2019efy}}) = \MpoleValCTNLO  \pm  \eMpolestat (\text{stat.})  \pm  \systexpTot (\text{syst.})\  \eMpoleScaleCTNLO \ (\text{scale})\ \eMpolePDFCTNLO\ (\text{PDF}\oplus\alphas) ~\gev,\\
& \Mpole (\text{MSHT20~\cite{Bailey:2020ooq}}) = \MpoleValMSHT  \pm  \eMpolestat (\text{stat.})  \pm  \systexpTot (\text{syst.})\  \eMpoleScaleMSHT\ (\text{scale})\    \eMpolePDFMSHT\ (\text{PDF}\oplus\alphas)  ~\gev,\\
& \Mpole (\text{NNPDF30~\cite{Ball:2014uwa}}) = \MpoleValNNPDF  \pm  \eMpolestat (\text{stat.})  \pm  \systexpTot (\text{syst.})   \ \eMpoleScaleNNPDF\ (\text{scale}) \   \eMpolePDFNNPDF\ (\text{PDF}\oplus\alphas) ~\gev, \\
& \Mpole (\text{ABMP16~\cite{Alekhin:2017kpj}}) = \MpoleValABMP + \eMpolestat (\text{stat.})   \pm \systexpTot (\text{syst.})\   \eMpoleScaleABMP \ (\text{scale}) \  \eMpolePDFABMP\ (\text{PDF}\oplus\alphas) ~\gev.\\
\end{split}
\label{eq:results_pdfs}
\end{equation}
The central values of the results obtained with CT18 and NNPDF30 agree within approximately 200~\MeV with each other and with the PDF4LHC21 result of Eq.~(\ref{eq:massresult}).

Several crosschecks are performed to verify the consistency of the results. The \Mpole mass is extracted from the $2\rightarrow7$ measurement with the calculation of Ref.~\cite{Bevilacqua:2016jfk}.
Despite the $2 \to 7$ cross-section measurement being slightly more precise than the $2 \to 3$ one, the smaller sensitivity close to the \ttbar{} threshold leads to an increase of all uncertainty components on the top-quark mass, yielding:
\begin{equation}
\topMassResultTotalOffshell \ .
\label{eq:massresultOffShell}
\end{equation}
The central value of the $2 \to 7$ result is approximately $1~\GeV$ higher than the $2 \to 3$ one. The difference between the two is covered by the scale uncertainty of the theoretical prediction, which grows from $0.3-0.4~\GeV$ in the $2 \to 3$ measurement to approximately 1~\GeV\ in the $2 \to 7$ one.
The result for the central value was cross-checked by repeating the $2 \to 7$ analysis with correction factors derived from the $bb4l$ MC simulation instead of the nominal correction factors, and obtaining the same value within  100~\MeV.%

The differential cross-section measurement and mass extraction are moreover repeated on sub-sets of the data, splitting the data by data-taking period. In each case, the data is corrected using the Monte Carlo simulation corresponding to the sub-set in question. Good agreement is found for the measured \Mpole\ values in different data-taking periods, with differences covered by single-period statistical uncertainties.

Finally, the top-quark mass measurement based on the $2\rightarrow3$ parton level prediction is repeated with different choices
for the $\pT$ cuts on the additional jet, at 30~\GeV~and 60~\GeV.
The \pT\ cut variation is synchronized at parton and detector level and the unfolding correction is re-derived for each value.
The results for \Mpole are found to be stable against the choice of the $\pT$ cut, with changes in the extracted mass value of $\pm250~\MeV$.
The total uncertainty of the measurement increases with the lower \pt\ requirement and the total uncertainty with the 60 GeV requirement is similar to the total uncertainty with the nominal selection.
Theoretical scale uncertainties for the 30~\GeV requirement are found to be 100~\MeV larger than for the 50~\GeV or 60~\GeV requirements.
The jet energy scale uncertainty is reduced by 150~\MeV when the selection requirement is increased,
while the statistical and some modelling uncertainties increase.
Future measurements with larger datasets may therefore benefit from using tighter selection requirements.
%
%
%
%

\begin{comment}
A number of cross-checks are performed to verify the consistency of the results obtained on sub-sets of the data, splitting the data by data-taking period, or decay channel. In each case, the data is corrected using the Monte Carlo simulation that corresponds to the sub-set in question.
The value of the top-quark pole mass was measured separately for the \ee, \emu, and \mumu\ channels, resulting in:
\begin{equation}
\begin{split}
&  \Mpole (\ee) = 169.0 \pm 0.8  (\text{stat.}) ~\gev \\
& \Mpole (\emu) = 169.9 \pm 0.4 (\text{stat.}) ~\gev \\
&  \Mpole (\mumu) = 169.0 \pm 0.5 (\text{stat.}) ~\gev, \\
\end{split}
\end{equation}
where only the statistical uncertainty is presented.
The results are compatible within the statistical uncertainties of the measurements.

The fit is also repeated on the three data-taking periods that compose run 2 of the LHC:
\begin{equation}
\begin{split}
& \Mpole (2015+16) = 170.7 \pm 0.5~\gev \\
& \Mpole (2017) = 168.8 \pm 0.5~\gev \\
& \Mpole (2018) = 168.9 \pm 0.5~\gev, \\
\end{split}
\end{equation}
%
%
%
%
%

Finally, the top-quark mass fit is repeated with different $\pT$ cuts on the additional jet, finding the following values:
\begin{equation}
\begin{split}
& \Mpole (p_T^{\text{extra}} > 30 \gev) = 169.6 \pm 0.2 ~\gev \\
& \Mpole (p_T^{\text{extra}} > 60 \gev) = 169.5 \pm 0.3 ~\gev, \\
\end{split}
\end{equation}
where the uncertainty is statistical. The results agree with each other and with the nominal result of Eq.~\ref{eq:massresult}.
\end{comment}

\begin{figure}[!ht]
\centering
\includegraphics[width=0.8\linewidth]{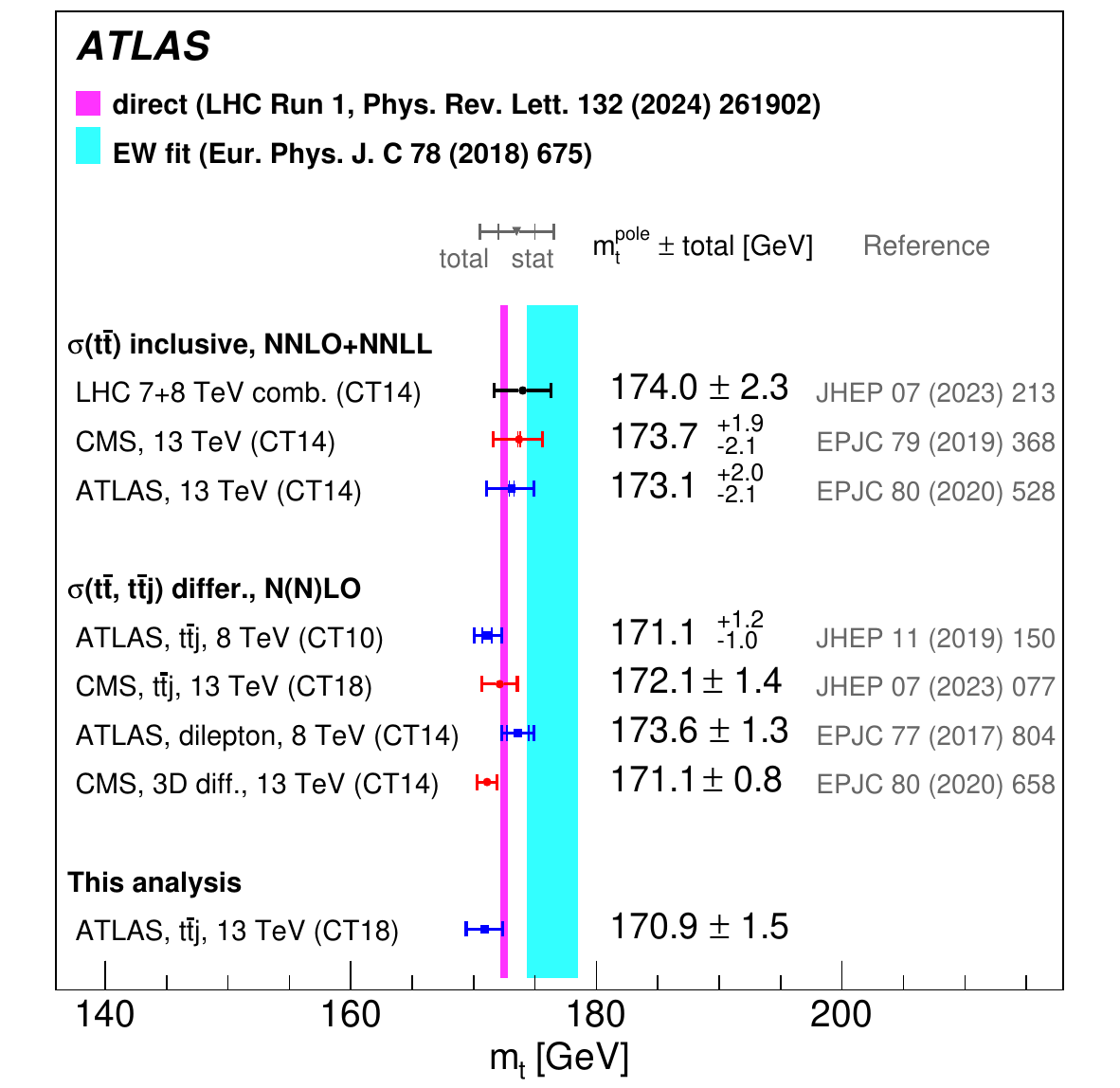}
\caption{Summary of top-quark mass results at the LHC, obtained from inclusive and differential top-quark-pair production cross-section measurements. For a more uniform comparison, results are chosen that were obtained with the most recent CTEQ PDFs at the time of the publication, as indicated in parentheses.  The values shown here do not correspond to the nominal result of the paper in all cases.  The value reported for this analysis is obtained from the $2\rightarrow3$ prediction.}
\label{fig:summary_pole_mass}
\end{figure}

The results of this analysis are compared with other top-quark mass determinations in Figure~\ref{fig:summary_pole_mass}.
The values of the top-quark pole mass in Eq.~(\ref{eq:massresult}) and Eq.~(\ref{eq:results_pdfs}) are in agreement with the determination by CMS using the same approach~\cite{CMS-TOP-21-008} and with earlier ATLAS results on Run~1 data~\cite{TOPQ-2014-06,TOPQ-2019-12}. The uncertainty of the result presented in this paper are slightly larger, mainly due to the more extensive set of modeling uncertainties considered.

These results agree within uncertainties with determinations of the top-quark pole mass from the inclusive cross-section~\cite{TOPQ-2018-39}. The central value of the latter tends to be somewhat higher: the ATLAS+CMS Run~1 combination~\cite{TOPQ-2018-39} yields \mbox{$\Mpole = $ 173.4 $^{+1.8}_{-2.0}$ \GeV{}} with a prediction based on a version of the NNPDF3.1 PDF without considering top-quark measurements as inputs to the PDF fit.
The values obtained in this analysis agree moreover with the precise determination of the top-quark mass obtained from the combination of ATLAS and CMS Run~1 direct measurements~\cite{ATLAS:2024dxp}, i.e. $m_{t} = 172.52\pm0.33~\GeV$.

The indirect estimate of the top-quark mass from a recent global electroweak fit~\cite{Haller2018} is also shown. The indirect determination of the top quark mass from electro-weak precision observables yields $m_t^\text{pole} = 176.4 \pm 2.1~\GeV$; a difference of about 5~\GeV from the nominal result of this analysis.

% End of text imported from the .//content/Results-MassValue.tex input file

%

%
%
\FloatBarrier

\section{Conclusion}
\label{sec:Conclusion}
%

% The next lines are included from the .//content/Conclusion.tex input file
A measurement is presented of the normalized differential cross-section of top-quark-antiquark pair production with an additional high-\pT\ jet %
as a function of $\Rhos = 2 m_0/\sqrt{s_{\ttbarjet~}}$, where $m_0=170$~\gev.
The \pp\ collision data set at $\rts=13$ \tev\ collected by the ATLAS experiment during Run 2 of the LHC is used, corresponding to an integrated luminosity of 140~\ifb.
The measurement is performed in the \emu~decay channel of the \ttbar\ system, where the neutrinos from the \Wboson-boson decays are reconstructed using a combination of the loose kinematic reconstruction and $\phi$-weighting method.
The \rhos\ distribution is corrected to parton level using the Iterative Bayesian Unfolding technique.
Two parton level definitions are employed, considering the top quarks as stable or not.
The normalized differential cross-section \Robs\ is presented in seven bins with a relative uncertainty ranging from a few \% to approximately 25~\% in the \rhos\ region with the largest mass sensitivity.

The top-quark pole mass \Mpole is extracted from the measured \Robs\ distribution with a least-squares fit to fixed-order calculations at NLO accuracy in the strong coupling. %
The top-quark pole mass obtained with the PDF4LHC21 PDF set, in the stable top-quark approximation, is
\begin{equation*}
m_{t}^\text{pole}=170.7\pm0.3~(\text{stat.})\pm1.4~(\text{syst.})~\pm 0.3~(\text{scale})~\pm 0.2~(\text{PDF}\oplus\alpha_\text{S})~\text{GeV}\quad ,
\end{equation*}
where the uncertainties represent the statistical uncertainty of the data, the experimental and modeling uncertainties from the differential cross-section measurement, the theoretical uncertainty estimated from the  choice of renormalization and factorization scales, and the impact of PDF and $\alpha_s$ variations on the theoretical prediction, respectively.
The top-quark pole mass is also determined based on cross-section calculations made with different PDF sets.

Several important cross-checks are performed to verify the stability of the result.
The result is found to be robust within $\pm$ 250~\MeV{} against variations of the \pT~cut on the additional jet in the interval between 30 and 60~\GeV. The result obtained with the $2 \to 3$ calculation assuming stable top quarks is compared with a result obtained from a fit of the corrected data to a $2 \to 7$ prediction where finite top-quark width and full off-shell effects are considered, hence testing the agreement in the description of top-quark decay modeling between the perturbative calculation and the Monte Carlo generator. %
The two results are found to agree within the theoretical uncertainties of the calculations.
The measured values of the top quark pole mass agrees, moreover, with the result of precise direct mass measurements, where the LHC Run~1 average $m_t = 172.52 \pm 0.33 $ \GeV{} is taken as a reference. Hence, this result provides a confirmation of the interpretation of direct mass measurements in terms of the pole mass scheme, within the uncertainty of the result. The top quark pole mass constitutes an important input to assess the stability of the electroweak vacuum and to fits testing the compatibility of electroweak precision measurements. The differential cross section data are publicly available and can be used in the future to repeat the mass extraction with improved calculations.

\clearpage

% End of text imported from the .//content/Conclusion.tex input file

%
\section*{Acknowledgements}
%

% The next lines are included from the .//acknowledgements/Acknowledgements.tex input file
%
%

%
%

We thank CERN for the very successful operation of the LHC and its injectors, as well as the support staff at
CERN and at our institutions worldwide without whom ATLAS could not be operated efficiently.

The crucial computing support from all WLCG partners is acknowledged gratefully, in particular from CERN, the ATLAS Tier-1 facilities at TRIUMF/SFU (Canada), NDGF (Denmark, Norway, Sweden), CC-IN2P3 (France), KIT/GridKA (Germany), INFN-CNAF (Italy), NL-T1 (Netherlands), PIC (Spain), RAL (UK) and BNL (USA), the Tier-2 facilities worldwide and large non-WLCG resource providers. Major contributors of computing resources are listed in Ref.~\cite{ATL-SOFT-PUB-2025-001}.

We gratefully acknowledge the support of ANPCyT, Argentina; YerPhI, Armenia; ARC, Australia; BMWFW and FWF, Austria; ANAS, Azerbaijan; CNPq and FAPESP, Brazil; NSERC, NRC and CFI, Canada; CERN; ANID, Chile; CAS, MOST and NSFC, China; Minciencias, Colombia; MEYS CR, Czech Republic; DNRF and DNSRC, Denmark; IN2P3-CNRS and CEA-DRF/IRFU, France; SRNSFG, Georgia; BMFTR, HGF and MPG, Germany; GSRI, Greece; RGC and Hong Kong SAR, China; ICHEP and Academy of Sciences and Humanities, Israel; INFN, Italy; MEXT and JSPS, Japan; CNRST, Morocco; NWO, Netherlands; RCN, Norway; MNiSW, Poland; FCT, Portugal; MNE/IFA, Romania; MSTDI, Serbia; MSSR, Slovakia; ARIS and MVZI, Slovenia; DSI/NRF, South Africa; MICIU/AEI, Spain; SRC and Wallenberg Foundation, Sweden; SERI, SNSF and Cantons of Bern and Geneva, Switzerland; NSTC, Taipei; TENMAK, T\"urkiye; STFC/UKRI, United Kingdom; DOE and NSF, United States of America.

Individual groups and members have received support from BCKDF, CANARIE, CRC and DRAC, Canada; CERN-CZ, FORTE and PRIMUS, Czech Republic; COST, ERC, ERDF, Horizon 2020, ICSC-NextGenerationEU and Marie Sk{\l}odowska-Curie Actions, European Union; Investissements d'Avenir Labex, Investissements d'Avenir Idex and ANR, France; DFG and AvH Foundation, Germany; Herakleitos, Thales and Aristeia programmes co-financed by EU-ESF and the Greek NSRF, Greece; BSF-NSF and MINERVA, Israel; NCN and NAWA, Poland; La Caixa Banking Foundation, CERCA Programme Generalitat de Catalunya and PROMETEO and GenT Programmes Generalitat Valenciana, Spain; G\"{o}ran Gustafssons Stiftelse, Sweden; The Royal Society and Leverhulme Trust, United Kingdom.

In addition, individual members wish to acknowledge support from CERN: European Organization for Nuclear Research (CERN DOCT); Chile: Agencia Nacional de Investigaci\'on y Desarrollo (FONDECYT 1230812, FONDECYT 1240864); China: Chinese Ministry of Science and Technology (MOST-2023YFA1605700, MOST-2023YFA1609300), National Natural Science Foundation of China (NSFC - 12175119, NSFC 12275265); Czech Republic: Czech Science Foundation (GACR - 24-11373S), Ministry of Education Youth and Sports (ERC-CZ-LL2327, FORTE CZ.02.01.01/00/22\_008/0004632), PRIMUS Research Programme (PRIMUS/21/SCI/017); EU: H2020 European Research Council (ERC - 101002463); European Union: European Research Council (BARD No. 101116429, ERC - 948254, ERC 101089007), European Regional Development Fund (SMASH COFUND 101081355, SLO ERDF), Horizon 2020 Framework Programme (MUCCA - CHIST-ERA-19-XAI-00), European Union, Future Artificial Intelligence Research (FAIR-NextGenerationEU PE00000013), Italian Center for High Performance Computing, Big Data and Quantum Computing (ICSC, NextGenerationEU); France: Agence Nationale de la Recherche (ANR-21-CE31-0022, ANR-22-EDIR-0002); Germany: Baden-Württemberg Stiftung (BW Stiftung-Postdoc Eliteprogramme), Bundesministerium für Bildung und Forschung (BMBF), Deutsche Forschungsgemeinschaft (DFG - 469666862, DFG - CR 312/5-2, DFG - 396021762 - TRR 257); China: Research Grants Council (GRF); Italy: Istituto Nazionale di Fisica Nucleare (ICSC, NextGenerationEU), Ministero dell'Università e della Ricerca (NextGenEU I53D23000820006 M4C2.1.1, SOE2024\_0000023); Japan: Japan Society for the Promotion of Science (JSPS KAKENHI JP22H01227, JSPS KAKENHI JP22H04944, JSPS KAKENHI JP22KK0227, JSPS KAKENHI JP24K23939, JSPS KAKENHI JP25H00650, JSPS KAKENHI JP25H01291, JSPS KAKENHI JP25K01023); México: Universidad Nacional Autónoma de México (DGAPA-PAPIIT IA102224); Norway: Research Council of Norway (RCN-314472); Poland: Ministry of Science and Higher Education (IDUB AGH, POB8, D4 no 9722), Polish National Science Centre (NCN 2021/42/E/ST2/00350, NCN OPUS 2023/51/B/ST2/02507, NCN OPUS nr 2022/47/B/ST2/03059, NCN UMO-2019/34/E/ST2/00393, UMO-2022/47/O/ST2/00148, UMO-2023/49/B/ST2/04085, UMO-2023/51/B/ST2/00920, UMO-2024/53/N/ST2/00869); Portugal: Foundation for Science and Technology (FCT); Spain: Ministry of Science and Innovation (MCIN \& NextGenEU PCI2022-135018-2, MICIN \& FEDER PID2021-125273NB, RYC2019-028510-I, RYC2020-030254-I, RYC2021-031273-I, RYC2022-038164-I); Sweden: Carl Trygger Foundation (Carl Trygger Foundation CTS 22:2312), Swedish Research Council (Swedish Research Council 2023-04654, VR 2021-03651, VR 2022-03845, VR 2022-04683, VR 2023-03403, VR 2024-05451), Knut and Alice Wallenberg Foundation (KAW 2018.0458, KAW 2022.0358, KAW 2023.0366); Switzerland: Swiss National Science Foundation (SNSF - PCEFP2\_194658); United Kingdom: Leverhulme Trust (Leverhulme Trust RPG-2020-004), Royal Society (NIF-R1-231091); United States of America: U.S. Department of Energy (ECA DE-AC02-76SF00515), Neubauer Family Foundation.

%
%

% End of text imported from the .//acknowledgements/Acknowledgements.tex input file

%
%
%
\printbibliography

\clearpage
% ATLAS Collaboration author list
% Reference date of TOPQ-2018-23 is 2025-05-20
% Author list last updated on date 02-JUL-25
% Data extracted on 02-Jul-2025 for paper reference TOPQ-2018-23
% at 12:04pm
 
\begin{flushleft}
\hypersetup{urlcolor=black}
{\Large The ATLAS Collaboration}

\bigskip

\AtlasOrcid[0000-0002-6665-4934]{G.~Aad}$^\textrm{\scriptsize 104}$,
\AtlasOrcid[0000-0001-7616-1554]{E.~Aakvaag}$^\textrm{\scriptsize 17}$,
\AtlasOrcid[0000-0002-5888-2734]{B.~Abbott}$^\textrm{\scriptsize 123}$,
\AtlasOrcid[0000-0002-0287-5869]{S.~Abdelhameed}$^\textrm{\scriptsize 119a}$,
\AtlasOrcid[0000-0002-1002-1652]{K.~Abeling}$^\textrm{\scriptsize 55}$,
\AtlasOrcid[0000-0001-5763-2760]{N.J.~Abicht}$^\textrm{\scriptsize 49}$,
\AtlasOrcid[0000-0002-8496-9294]{S.H.~Abidi}$^\textrm{\scriptsize 30}$,
\AtlasOrcid[0009-0003-6578-220X]{M.~Aboelela}$^\textrm{\scriptsize 45}$,
\AtlasOrcid[0000-0002-9987-2292]{A.~Aboulhorma}$^\textrm{\scriptsize 36e}$,
\AtlasOrcid[0000-0001-5329-6640]{H.~Abramowicz}$^\textrm{\scriptsize 157}$
% ...
\bigskip
\\

% ...
$^{17}$Department for Physics and Technology, University of Bergen, Bergen; Norway.\\
% ...
$^{30}$Physics Department, Brookhaven National Laboratory, Upton NY; United States of America.\\
% ...
$^{36}$$^{(a)}$Facult\'e des Sciences Ain Chock, Universit\'e Hassan II de Casablanca;$^{(b)}$Facult\'{e} des Sciences, Universit\'{e} Ibn-Tofail, K\'{e}nitra;$^{(c)}$Facult\'e des Sciences Semlalia, Universit\'e Cadi Ayyad, LPHEA-Marrakech;$^{(d)}$LPMR, Facult\'e des Sciences, Universit\'e Mohamed Premier, Oujda;$^{(e)}$Facult\'e des sciences, Universit\'e Mohammed V, Rabat;$^{(f)}$Institute of Applied Physics, Mohammed VI Polytechnic University, Ben Guerir; Morocco.\\
% ...
$^{45}$Physics Department, Southern Methodist University, Dallas TX; United States of America.\\
% ...
$^{49}$Fakult\"{a}t Physik , Technische Universit{\"a}t Dortmund, Dortmund; Germany.\\
% ...
$^{55}$II. Physikalisches Institut, Georg-August-Universit\"{a}t G\"ottingen, G\"ottingen; Germany.\\
% ...
$^{104}$CPPM, Aix-Marseille Universit\'e, CNRS/IN2P3, Marseille; France.\\
% ...
$^{119}$$^{(a)}$New York University Abu Dhabi, Abu Dhabi;$^{(b)}$United Arab Emirates University, Al Ain; United Arab Emirates.\\
% ...
$^{123}$Homer L. Dodge Department of Physics and Astronomy, University of Oklahoma, Norman OK; United States of America.\\
% ...
$^{157}$Raymond and Beverly Sackler School of Physics and Astronomy, Tel Aviv University, Tel Aviv; Israel.\\
% ...
$^{a}$ Also at Affiliated with an institute formerly covered by a cooperation agreement with CERN.\\
$^{b}$ Also at An-Najah National University, Nablus; Palestine.\\
$^{c}$ Also at Borough of Manhattan Community College, City University of New York, New York NY; United States of America.\\
$^{d}$ Also at Center for Interdisciplinary Research and Innovation (CIRI-AUTH), Thessaloniki; Greece.\\
$^{e}$ Also at Centre of Physics of the Universities of Minho and Porto (CF-UM-UP); Portugal.\\
$^{f}$ Also at CERN, Geneva; Switzerland.\\
% ...

\end{flushleft}

% Created with Glance <Atlas.Glance@cern.ch>

%
%
%
\clearpage

\end{document}